\documentclass[twocolumn,secnumarabic,amssymb, nobibnotes, aps, pre]{revtex4-2}

\usepackage{framed}
\usepackage{amsfonts}
\usepackage{amsmath}
\usepackage{graphicx}
\usepackage{color}

\usepackage[hidelinks]{hyperref}
\usepackage{xcolor}
\hypersetup{
	colorlinks,
	linkcolor={blue!80},
	citecolor={blue},
	urlcolor={blue!80!black}
}

\begin{document}
\title{Dynamics of three-dimensional spatiotemporal solitons in multimode waveguides}

\author{Pedro Parra-Rivas}
\email{pedro.parra-rivas@uniroma1.it}
\author{Yifan Sun, and Stefan Wabnitz }
\affiliation{
	Dipartimento  di  Ingegneria  dell’Informazione, Elettronica  e  Telecomunicazioni,
	Sapienza  Universit{\'a}  di  Roma, via  Eudossiana  18, 00184  Rome, Italy\\
}

\begin{abstract}
	In this work, we present a detailed study of the dynamics and stability of fundamental spatiotemporal solitons emerging in multimode waveguides with a parabolic transverse profile of the linear refractive index. Pulsed beam propagation in these structures can be described by using a Gross-Pitaevskii equation with a two-dimensional parabolic spatial potential. Our investigations are based on comparing variational approaches, based on the Ritz optimization method, with extensive numerical simulations. We found that, with a Kerr self-focusing nonlinearity, spatiotemporal solitons are stable for low pulse energies, where our analytical results find a perfect agreement with the numerical simulations. However, solitons with progressively increasing energies eventually undergo a wave collapse, which is not predicted within the variational framework. In a self-defocusing scenario, again for low energies there is good agreement between the variational predictions and simulations. Whereas, for large soliton energies complex spatiotemporal dynamics emerge. 
\end{abstract}
\maketitle


\section{Introduction}

{\it Solitons} or {\it solitary waves} are localized nonlinear wave packets which propagate without suffering any shape modification. They arise from the interplay between linear and nonlinear processes, which separately would cause the wave to decay. These localized waves were described, for the first time, in 1834 by John Scott Russell, who observed the propagation of a solitary wave with these characteristics in the union canal in Scotland. Russell named it a {\it wave of translation} \cite{dauxois_physics_2006}. However, the word soliton was coined much later, in 1965, by Zabuski and Kruskal while studying pulse interactions in collisionless plasmas \cite{PhysRevLett.15.240}. Since then, solitons have been discovered and studied in a large variety of physical contexts, including hydrodynamics, plasmas, and condensed matter physics, biology, and nonlinear optics, to cite a few \cite{dauxois_physics_2006,kartashov_frontiers_2019,malomed_multidimensional_nodate}.

In the realm of nonlinear optics, these particle-like objects may emerge in nonlinear media owing to a balance between dispersive or diffractive effects and light confinement in either time or space, leading to temporal or spatial solitons, respectively \cite{kivshar_optical_2003}. 
Light confinement may be related to the intensity-dependent contribution to the refractive index of the material (optical Kerr effect), which yields spatial self-(de)focusing or temporal self-phase modulation in each of the previous two scenarios. In nonlinear dispersive media, such as singlemode optical fibers, temporal broadening of a light wavepacket (induced by chromatic dispersion) can be counteracted by self-phase modulation, leading to the formation of {\it temporal solitons}, which propagate unchanged in the longitudinal direction. In contrast to these states, {\it spatial solitons} form transversally to the direction of beam propagation, whenever the  
natural diffraction-induced spreading of a light beam is compensated for by the spatial (nonlinear) self-focusing effect. Thus, in each case, a balance between a spatial coupling mechanism (i.e., diffraction or dispersion) and nonlinearity is a necessary condition for soliton formation \cite{kivshar_optical_2003}. 

More complex is the situation where the previous
temporal and spatial effects couple simultaneously. In this case, dispersion and diffraction may counteract nonlinearity at once, leading to light confinement in space-time, and therefore to the formation of a large variety of coherent three-dimensional spatiotemporal states, including fundamental solitons and vortex states \cite{kartashov_frontiers_2019}. 

Fundamental {\it spatiotemporal solitons} (STS), named also hereafter {\it light bullets} according to Silberberg's terminology \cite{silberberg_collapse_1990}, do not carry vorticity (i.e., they are spinning less), and are generally affected by various propagation instabilities such as {\it spatiotemporal wave collapse} \cite{silberberg_collapse_1990,berge_wave_1998,bang_collapse_2002}, which make them challenging to observe. Wave collapse occurs whenever a strong contraction or compression of a nonlinear wave leads to a catastrophic blowup of its amplitude after a finite time or propagation distance \cite{berge_wave_1998,bang_collapse_2002}. The contraction suffered by the wave needs two or more dimensions, in order to be strong enough to generate the collapse; therefore, wave collapse is absent in 1D geometries. This fundamental phenomenon not only arises in nonlinear optics \cite{garmire_dynamics_1966}, but it also appears in different nonlinear wave contexts, ranging from Bose-Einstein condensates (BECs) to astrophysics \cite{sackett_measurements_1999,wong_three-dimensional_1984,noauthor_black_1983}. Hence, a central challenge in the scientific community is to find robust mechanisms, which may be able to arrest these destructive wave phenomena \cite{kartashov_frontiers_2019,malomed_multidimensional_2016}. 

In this work, we study one of such mechanisms, leading to the stabilization of STSs in single-pass optical waveguides: it is associated with the presence of a radially symmetric, parabolic refractive index profile in the transverse plane, perpendicular to the light propagation direction \cite{yu_spatio-temporal_1995,raghavan_spatiotemporal_2000}.
This material inhomogeneity appears naturally in graded-index (GRIN) waveguides such as multimode (MM) fibers \cite{horak_multimode_2012}. Here, the parabolic spatial profile of the linear refractive index acts as spatial guiding potential, and it is able to arrest spatiotemporal wave collapse, as experimentally demonstrated in Ref. ~\cite{renninger_optical_2013}.
However, in most cases, the formation of 3D solitons is elusive and they have been only observed as a transient phenomenon \cite{panagiotopoulos_super_2015,minardi_three-dimensional_2010}.

Based on Lagrangian and Hamiltonian variational approaches and advanced numerical simulations, we perform a detailed characterization of the dynamics and stability of the fundamental STSs which emerge in this system, under different regimes of operation. Our investigations go beyond those presented in \cite{yu_spatio-temporal_1995,raghavan_spatiotemporal_2000,shtyrina_coexistence_2018}, and demonstrate the degree of agreement between analytical approximations and direct numerical solutions over a wide range of STSs energies, which was not previously explored. 
Specifically, we find that there is a perfect agreement between the variational approach and numerical simulations for low STS energies. However, such agreement worsens when the STS energy increases. Indeed, for large enough STS energies, the bullets may undergo spatiotemporal wave collapse and exhibit other complex dynamics, which so far have remained unexplored.

The paper is organized as follows. In Section~\ref{sec:1} we introduce our model, and its associated variational formulations in terms of Lagrangian and Hamiltonian densities, respectively. Section~\ref{sec:2} contains a general introduction to the Ritz optimization method in terms of the Lagrangian and Hamiltonian formalism. In Section~\ref{sec:3} we apply this method to the case of shape-preserving (i.e., steady state) fundamental STSs. In Sections~\ref{sec:4}, by using the Lagrangian formalism, we extend the Ritz method in order to capture the $z$-dependence of the STSs propagation. By doing so, we are able to reduce the initial infinite-dimensional model to a finite-dimensional (effective) dynamical system. A similar system is then obtained in Section~\ref{sec:5} by using, this time, a Hamiltonian approach. Later, in Section~\ref{sec:6} we analyze the STS stability by using different stability criteria, including the Vakhitov-Kolokolov and Lyapunov criterion. After this, in Section~\ref{sec:7} we test our analytical results by performing full 3D numerical simulations of the original model. Finally, in Section~\ref{sec:8} we present a short discussion, draw our conclusions and comment on future research directions. 

\section{Variational formulation of the Gross-Pitaevskii equation with a 2D parabolic potential}\label{sec:1}
The scalar electric field of an optical wave propagating in a MM waveguide can be described in terms of the dimensionless 3D+1 Gross-Pitaevskii equation (GPE) as follows \cite{horak_multimode_2012}
\begin{equation}\label{GPE}
\partial_z u=\frac{i}{2}\nabla_\perp^2u+i\frac{\delta}{2}\partial^2_t u+i\frac{\rho}{2}(x^2+y^2)u+i\nu|u|^2u.
\end{equation}
Here $u=u(x,y,t,z)$ is the normalized electric field component of the wave propagating along the $z$-direction, $\nabla_\perp^2\equiv\partial_x^2+\partial_y^2$ represents diffraction, $\partial_t^2$ represents chromatic or group velocity dispersion (GVD), with the coefficient  $\delta=\pm 1$ for the anomalous/normal dispersion regime, respectively, $\nu=\pm 1$ for self-focusing/self-defocusing Kerr nonlinearity, and $(x^2+y^2)$ is the 2D parabolic potential describing the transverse spatial profile of the linear refractive index of the material \cite{horak_multimode_2012}. Here, $\rho=-1$ ($\rho=1$) is chosen for guiding (antiguiding) materials.

This same equation can be used in the context of BECs, in order to describe nearly 1D condensates, with a cigar-shaped trapping potential ($\rho<0$) if we exchange the $z$ coordinate with $t$ \cite{strecker_formation_2002,malomed_multidimensional_2016}. In this context, $\nu=1$ models a self-attractive nonlinearity \cite{malomed_multidimensional_nodate}.  

Equation~(\ref{GPE}) possesses the Lagrangian density 
\begin{equation}
\begin{aligned} 
	\mathcal{L}=&-\frac{1}{2}\left(|u_x|^2+|u_y|^2\right)-\frac{\delta}{2}|u_t|^2+\frac{\rho}{2} (x^2+y^2)|u|^2\\
	&+\frac{\nu}{2}  |u|^4 +\frac{i}{2} \left(u^*u_z-u u_z^* \right),
\end{aligned}
\label{Lagran1}
\end{equation}
where we have rewritten the derivatives as $u_\xi\equiv\partial_\xi u$, with $\xi$ being any variable $x,y,z$ and $t$. This Lagrangian density contains all relevant information about the system dynamics, including its conservation laws \cite{abraham_foundations_2008}. Indeed, from the Lagrangian density one recovers Eq.~(\ref{GPE}) from the Euler-Lagrange equations \cite{wiggins_introduction_2003,abraham_foundations_2008}
\begin{equation}
\frac{d}{dz}\left(\frac{\partial\mathcal{L}}{\partial u^*_z}\right)+\frac{d}{dt}\left(\frac{\partial\mathcal{L}}{\partial u^*_t}\right)+\frac{d}{dx}\left(\frac{\partial\mathcal{L}}{\partial u^*_x}\right)+\frac{d}{dy}\left(\frac{\partial\mathcal{L}}{\partial u^*_y}\right)-\frac{\partial \mathcal{L}}{\partial u^*}=0.
\end{equation}
By defining the generalized field momenta $\mathcal{P}\equiv \partial_{u^*_z}\mathcal{L}=-iu/2$ and $\mathcal{P}^*\equiv \partial_{u_z}\mathcal{L}=iu^*/2$, our system can be described by using the Hamiltonian density, which is obtained from the Legendre transform \cite{wiggins_introduction_2003,abraham_foundations_2008}
\begin{equation}
\mathcal{H}=\mathcal{P} u^*_z+\mathcal{P}^* u_z-\mathcal{L}.   
\end{equation}
This transformation leads to 
\begin{equation}\label{Hamil_density}
\mathcal{H}=\frac{1}{2}\left(| u_x|^2+|u_y|^2\right)+\frac{\delta}{2}|u_t|^2-\frac{\nu}{2}  |u|^4 -\frac{\rho}{2} \left(x^2+y^2\right)|u|^2. 
\end{equation}
Equation~(\ref{GPE}) can also be derived from the Hamiltonian function through the Hamiltonian field equations \cite{abraham_foundations_2008}.

Shape-preserving (i.e., steady) spatiotemporal states can be written in the form $u(x,y,z,t)=v(x,y,t)e^{i\mu z}$, 
where $\mu$ is the propagation constant (or chemical potential in the context of BECs), and $v(x,y,t)$ is a real-valued function, describing the steady-state field. When applied to Eq.~(\ref{GPE}), this transformation leads to the $z$-independent (real) partial differential equation  
\begin{equation}\label{real_GP}
\frac{1}{2}\nabla_\perp^2v+\frac{\delta}{2} v_t+\frac{\rho}{2} (x^2+y^2)v+\nu v^3-	\mu v=0.
\end{equation}
Similarly to the previous case, the $z$-independent Eq.~(\ref{real_GP}) can be obtained from the Euler-Lagrange equations
\begin{equation}
\frac{d}{dt}\left(\frac{\partial\mathcal{L}_s}{\partial v_t}\right)+\frac{d}{dx}\left(\frac{\partial\mathcal{L}_s}{\partial v_x}\right)+\frac{d}{dy}\left(\frac{\partial\mathcal{L}_s}{\partial v_y}\right)-\frac{\partial \mathcal{L}_s}{\partial v}=0,
\end{equation}
where the stationary Lagrangian density is now defined as follows
\begin{equation}\label{static_Lagran_den}
\mathcal{L}_s\equiv-\frac{\delta}{4}v_t^2-\frac{1}{4}\left(v_x^2+v_y^2\right)+\frac{\rho}{4}(x^2+y^2)v^2+\frac{\nu}{4}v^4-\frac{\mu}{2}v^2.
\end{equation}
Note that this new Lagrangian depends explicitly on $\mu$, which is different from the Lagrangian density that we have previously defined in Eq.~(\ref{Lagran1}).

\section{The Ritz optimization method}\label{sec:2}
In this section we introduce a variational method that is widely used in order to compute soliton solutions, such as fundamental STSs or multidimensional solitons, in non-integrable conservative systems:
the Ritz optimization method \cite{anderson_selftrapped_1979,bondeson_soliton_1979, perez-garcia_dynamics_1997,malomed_variational_2002}. This method allows us to compute approximate analytical solutions of a given nonlinear partial differential equation by applying the principle of least action to a parameter-dependent solution ansatz, based on either the Lagrangian or the Hamiltonian formalism. There are other methods, such as the moment approach \cite{rasmussen_blow-up_1986,montesinos_stabilization_2004,hansson_nonlinear_2020}, that could be used to compute approximate analytical solutions for this type of equation, leading to similar results. However, these methods will not be considered in the present work. 

In our context, this Ritz variational approach was used before, in order to predict the existence of 3D STSs in inhomogeneous Kerr nonlinear media \cite{yu_spatio-temporal_1995,raghavan_spatiotemporal_2000}. In what follows, we describe the main steps in either the Lagrangian or the Hamiltonian formalism. 
\subsection{The Ritz method in the Lagrangian formalism}
The method consists of the following four main steps:
\begin{itemize}
\item[(1)]  First of all, we need to define an approximate ansatz solution, or trial function, that captures the main features and shape of the state that we want to compute. This ansatz is a function of the form $u=u[x,y,t;q(z)]$, which depends on $z$ through a number of parameters $$q(z)=\{q_1(z),\cdots,q_n(z)\},$$ which are the generalized coordinates of the system.

\item[(2)] Next, we calculate the effective Lagrangian function of the system, which is defined as 
\begin{equation} 
	L[q(z)]\equiv\int_{{\rm I\!R}^3} \mathcal{L}\left(u,u_t^2,\nabla_\perp^2;u[x,y,t,q(z)]\right)dxdydt.
\end{equation}

\item[(3)] After that, we obtain the dynamical system for the generalized coordinates by computing the Euler-Lagrange equations 
\begin{equation}\label{red_effec}
	\frac{d}{dz}\left(\frac{\partial L}{\partial (d_z q_m)}\right)-\frac{\partial L}{\partial q_m}=0,
\end{equation}
for each parameter $q_m$ with $m=1,\cdots,n$, where we have defined $d_z q_m\equiv dq_m/dz$. 

\item[(4)] Finally, in the last step, we study the dynamics of the reduced system (\ref{red_effec}), and rebuild the dynamics of the STS by using the initial ansatz.
\end{itemize}
\subsection{The Ritz method in the Hamiltonian formalism}
Equivalently, one may consider the Hamiltonian formalism for obtaining an effective reduced system. The process is equivalent to that followed in the Lagrangian case, but now we use the  Hamiltonian 
\begin{equation}\label{efec_hamilton}
H[q(z),p(z)]\equiv\int_{{\rm I\!R}^3}  \mathcal{H}\left(u,u_t^2,\nabla_\perp^2;u[x,y,t,q(z)]\right)dxdydt,
\end{equation}
where $p(z)=\{p_1(z),\cdots,p_n(z)\}$ are the generalized momenta defined as $$p_m=\frac{\partial L}{\partial (d_z q_m)}.$$

Then, the effective dynamics of the system are captured by the Hamiltonian equation of motion 
\begin{equation}
\frac{d q_m}{d z}=\frac{\partial H}{\partial p_m},\qquad
\frac{d p_m}{d z}=-\frac{\partial H}{\partial q_m},
\end{equation}
for each $m=1,\cdots,n$. After studying this reduced system, we rebuild the bullet behavior by using the ansatz solution (\ref{ansatz1}).

\section{Shape preserving spatiotemporal solitons}\label{sec:3}
In this section, we will compute an approximate solution for a shape-preserving (i.e., steady state) STSs, by applying the Ritz optimization method to the stationary Lagrangian density (\ref{static_Lagran_den}). 
Although this approach does not give us any information about the transient behavior or propagation dynamics, it allows us to compute the  propagation constant $\mu$ and, therefore, to estimate the stability of the STSs through the Vakhitov-Kolokolov criterium \cite{vakhitov_stationary_1973}, as it has been demonstrated in previous papers \cite{sakaguchi_two-dimensional_2006,malomed_stability_2007,desyatnikov_three-dimensional_2000,baizakov_multidimensional_2004}. We will come back to this stability analysis in Sec.~\ref{sec:6.2}. 

At this stage, it is essential to define a proper {\it trial function} or solution ansatz for analytically describing the STSs. The selection of this ansatz is not entirely arbitrary, but it is justified by different preliminary observations, mostly related to the symmetries of the system. For example, in the absence of a 2D parabolic potential (of with radially symmetric potentials), Eq.~(\ref{GPE}) may have radially symmetric 3D solitons, whose shape can be captured by just considering the radius $r^2=t^2+x^2+y^2$ as the only variable \cite{desyatnikov_three-dimensional_2000,desaix_variational_1991,skarka_spatiotemporal_1997,bang_collapse_2002,skarka_stability_2006}. In our current case, however, the potential is 2D, and the $r$-dependent ansatz is not valid. In this case, we choose the steady-state STS ansatz 
\begin{equation}\label{ansatz1}
v(x,y,t;\eta,a,A)= A{\rm sech}(\eta t) {\rm Exp}\left(-\frac{x^2+y^2}{2a^2}\right), 
\end{equation}
where $a>0$ is the width of the spatial Gaussian profile, $\eta>0$ is the inverse of the temporal width, and $A>0$ is the amplitude of the pulse. The justification of this choice is based on two main observations: (a) in the absence of dispersion and nonlinearity, Eq.~(\ref{GPE}) has Laguerre-Gaussian mode solutions, and the fundamental mode is a Gaussian; (b) in the absence of diffraction and spatial potential, Eq.~(\ref{GPE}) possesses a sech-shape bright soliton solution in the anomalous GVD regime \cite{kivshar_optical_2003}.

Note that this problem can also be analyzed by just considering a Gaussian ansatz, which makes the calculations simpler. This approach was followed in
\cite{shtyrina_coexistence_2018}.

By using the definition of the pulse energy  
\begin{eqnarray*}
E\equiv\int_{{\rm I\!R}^3} |u(x,y,t)|^2dxdydt=\int_{{\rm I\!R}^3} v(x,y,t)^2dxdydt,
\end{eqnarray*}
we obtain that
\begin{eqnarray*}
A=\sqrt{\frac{\eta E}{2\pi a^2}},
\end{eqnarray*}
and we can make our ansatz [i.e., Eq.~(\ref{ansatz1})] energy-dependent. In this way, the pulse energy becomes the most important control parameter for the STS solutions. Thus, we have that $q=\{q_1,q_2,q_3\}=\{\eta,a,E\}$.

\begin{figure*}[tbp]
\centering
\includegraphics[scale=1]{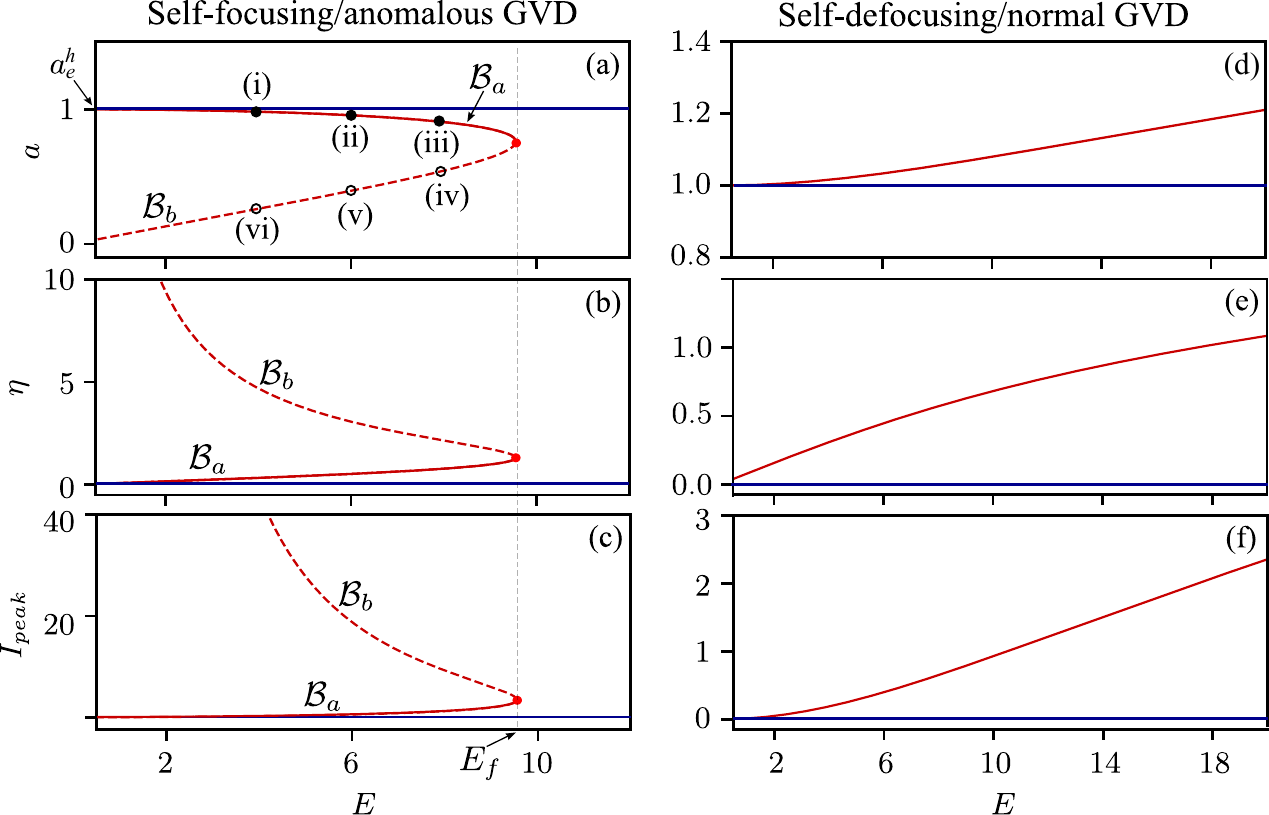}
\caption{Bifurcation diagrams for STS states as a function of $E$. Left column: Self-focusing/anomalous GVD. (a) Panel shows the  width of the STS as a function of $E$; (b) Panel shows the inverse of the temporal width $\eta$, or (c) the STS peak intensity $I_{peak}$.
	The branch of solutions $\mathcal{B}_a$ is plotted in solid, while $\mathcal{B}_b$ uses a dashed line. Labels (i)-(vi) correspond to the STSs depicted in Fig.~\ref{fig2}. Right column: Self-defocusing and normal GVD regime.
	From (d)-(f) we plot $a,\eta,I_{peak}$ and $\mu$ as a function of $E$. 
}
\label{fig1}    
\end{figure*}
With this ansatz, the static Lagrangian 
\begin{equation}
L_s(q)=\int_{{\rm I\!R}^3} \mathcal{L}_s[u,u_t^2,\nabla_\perp^2;u(q)]dxdydt
\end{equation}
reduces to
\begin{equation}
L_s=\frac{E}{12}\left(\frac{1}{a^2}\left(\frac{E\eta\nu}{2\pi}-3\right)-\delta \eta^2-6\mu+3\rho a^2\right).
\end{equation}
In this case, the effective Euler-Lagrange equations (\ref{red_effec}) become
\begin{equation}
\frac{\partial L_s}{\partial \eta}=0, \qquad \frac{\partial L_s}{\partial a}=0, \qquad \frac{\partial L_s}{\partial E}=0,
\end{equation}
which respectively lead to the set of equations
\begin{equation}\label{fixed_1}
\frac{E\nu}{a^2}-4\pi\delta\eta=0,  
\end{equation}
\begin{equation}\label{fixed_2}
E\eta\nu-6\pi(1+\rho a^4)=0, 
\end{equation}
\begin{equation}\label{fixed_3}
\mu=-\frac{1}{2a^2}\left(1-\frac{E\eta\nu}{3\pi}\right)-\frac{\delta\eta^2a^2}{3}+\rho a^4,
\end{equation}
providing that $a>0$.
Combining these expressions, one finally obtains that the steady-state soliton parameters satisfy 
\begin{equation}\label{eq1_static}
E=2\pi a\sqrt{6\delta(1+\rho a^4)} 
\end{equation}
\begin{equation}\label{eq2_static}
\eta=\frac{E\nu}{4\pi\delta a^2}=\frac{\nu}{2\delta a}\sqrt{6\delta(1+\rho a^4)} ,
\end{equation} 
whereas the propagation constant reads as
\begin{equation}
\mu=-\frac{1}{2}\left(1+\frac{1}{a^2}\right)+\frac{1}{a^2}\left(1+\rho a^4\right)+\frac{1}{2}\rho a^4
\end{equation}
Note that all of the previous quantities are parameterized by the spatial width coefficient $a$.


Moreover, the soliton parameters allow us to compute the peak soliton intensity (i.e., the intensity at the center of the bullet) as
\begin{equation}\label{Eq_inten}
I_{peak}=|A|^2=\frac{E\eta}{2\pi a^2}=\frac{3}{\nu a^2}(1+\rho a^4).
\end{equation}

At this stage, we can already obtain some general insights about our system. From Eqs.~(\ref{eq1_static}) and (\ref{eq2_static}) we find that, in order to obtain real solutions, it is required that $\delta(1+\rho a^4)>0$. When $1+\rho a^4=0$ (i.e., if $a^4=-1/\rho$), $E$ and $\eta$ becomes zero. This means that, with $E\rightarrow 0$, the temporal width of the state $\eta^{-1}\rightarrow \infty$, and the STS becomes a continuous-wave (CW) state of the system, which is homogeneous in time.  Furthermore, from Eq.~(\ref{eq2_static}) we see that $\delta$ and $\nu$ must have the same sign, in order for $\eta$ to be positive. 

Equation (\ref{eq1_static}) can also be written in the form 
\begin{equation}\label{Eq_for_a}
\rho a^6+a^2-\frac{1}{6\delta}\left(\frac{E}{2\pi}\right)^2=0,
\end{equation}
and it may have either one or two positive real roots, depending on the signs of $\nu$, $\delta$, and $\rho$. Unfortunately, this equation does not possess exact analytical solutions, and we need to solve it either by using approximate analytical methods, or numerically. 
\begin{figure*}[!t]
\centering
\includegraphics[scale=0.9]{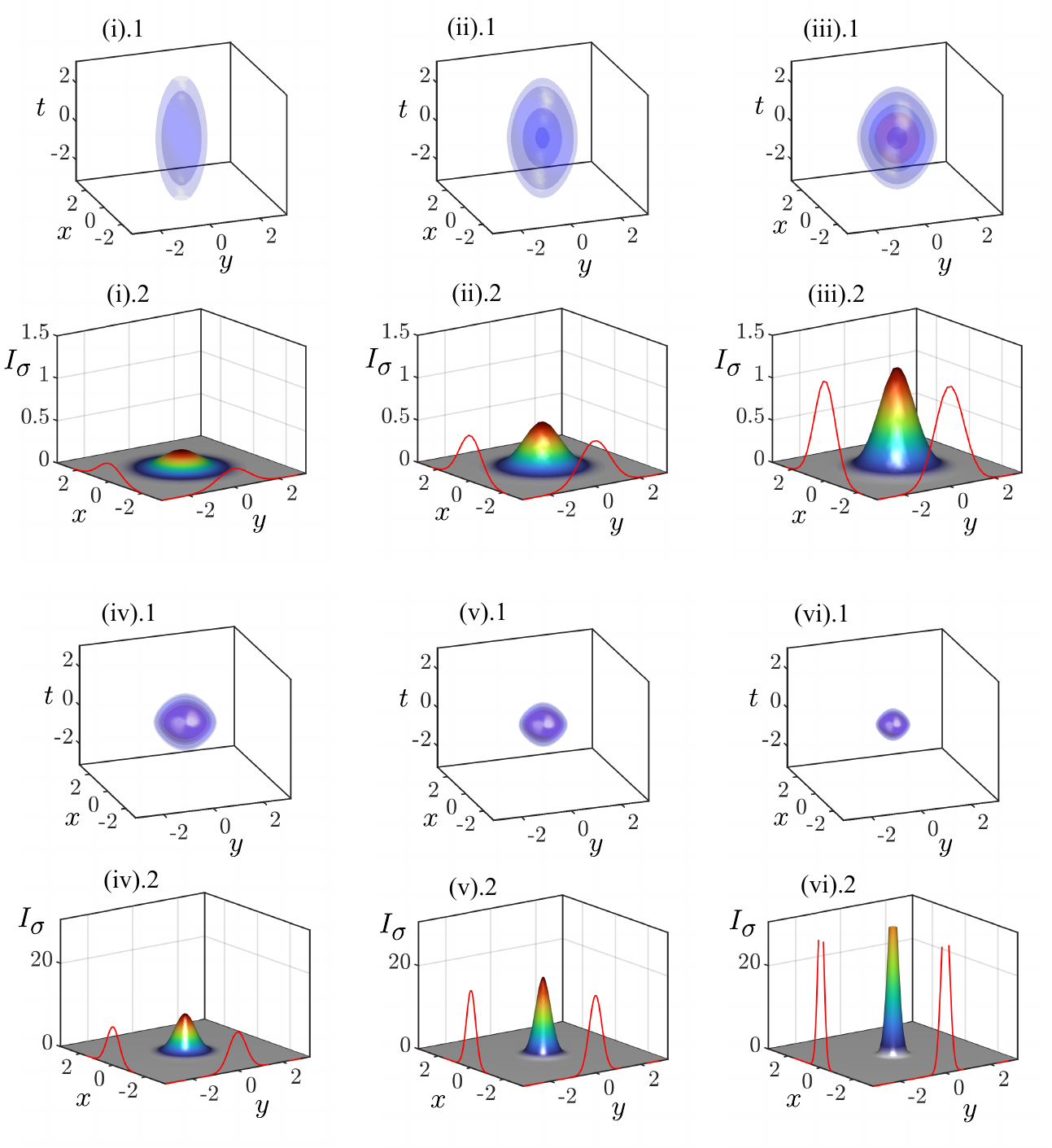}
\caption{STS states in the self-focusing/anomalous GVD regime. Panels (i).1-(iii).1 show the $\mathcal{B}_a$-STSs reconstruction corresponding to labels (i)-(iii) in Fig.~\ref{fig1}(a) for $E=4,6$ and $8$, respectively. Here we have plotted five isosurfaces at different peak intensities, namely $I=0.08,0.12.0.3,0.5,1.0$. Panels (i).2-(iii).2 illustrate the $t=0$ cross-ection intensity $I_\sigma\equiv I(x,y,t=0)$ for the STSs shown above. Panels (iv)-(vi) show analogous information than (i)-(iii), but for $\mathcal{B}_b$-related states. 
}
\label{fig2}    
\end{figure*}
In what follows, we will consider {\it guiding media}, and therefore we shall take $\rho<0$.

Depending on the signs of $\nu$ and $\delta$, we may consider the two main scenarios:
\begin{itemize}
\item $\nu=\delta=1$: {\it Self-focusing material with anomalous GVD},
\item $\nu=\delta=-1$: {\it Self-defocusing material with normal GVD}.
\end{itemize}
In what follows, we will analyze the features of these two scenarios separately. 

\begin{table}
\centering
\begin{tabular}{cccccc}
	
	$\mathcal{B}$&Label& $E$ & $a$ & $\eta$ & $I_{peak}$\\
	\hline\hline
	$\mathcal{B}_a$&(i)	& 4			&0.982009 & 0.33008&0.217906 \\
	\hline
	$\mathcal{B}_a$&(ii)& 6& 0.9555& 0.522974&	0.547004		\\
	\hline
	$\mathcal{B}_a$	&(iii)& 8&0.904689 &0.777824 &	1.21002		\\
	\hline
	$\mathcal{B}_b$&(iv)&8&0.544214&2.14952&9.24086\\
	\hline
	$\mathcal{B}_b$&(v)&6&0.394665&3.06537&18.793\\
	\hline
	$\mathcal{B}_b$&(vi)&4&0.260499&4.69069&44.0051\\
	\hline
\end{tabular}
\caption{Fixed points associated with the STSs shown in Fig.~\ref{fig2} and Fig.~\ref{fig1}(a), corresponding to the anomalous GVD/self-focusing nonlinearity.}
\label{table1}
\end{table}

\subsection{Guiding, self-focusing material with anomalous GVD}\label{sec: anomalous}
In the case of a guiding, self-focusing nonlinear material ($\nu=1$) with anomalous GVD  ($\delta=1$), Eqs.~(\ref{eq1_static}), 
(\ref{eq2_static}), and (\ref{Eq_inten})
become
\begin{eqnarray}
E=2\pi a\sqrt{6(1+\rho a^4)}, 
\end{eqnarray}
\begin{eqnarray}
\eta=\frac{1}{2 a}\sqrt{6(1+\rho a^4)}, 
\end{eqnarray}
\begin{eqnarray}
I_{peak}=\frac{3}{a^2}(1+\rho a^4).
\end{eqnarray}
Figures~\ref{fig1}(a)-(c) show the modification of these quantities as a function of $E$ for $\rho=-1$. In this regime, Eq.~(\ref{Eq_for_a}) has, for a fixed value of $E$, two real solutions, which correspond to the solution branches $\mathcal{B}_a$ (solid red line) and $\mathcal{B}_b$ (dashed red line), respectively. These two solution branches coexist between $E=E_0\equiv0$ and the fold, or turning point, taking place at $E=E_f$ (see the blue dot in Fig.~\ref{fig1}).
The fold position can be calculated analytically, by solving the equation $dE/da=0$, which leads to 
$1+3\rho a^4=0,$
providing that $1+\rho a^4>0$. The solution of this equation yields the fold parameters 
\begin{equation}
a_f=(-3\rho)^{-1/4},\qquad \eta_f=(-3\rho)^{1/4}, \qquad 
\end{equation}
\begin{equation}
E_f=4\pi a_f, \qquad I_{f}=\frac{2}{a_f^2}.
\end{equation}
This scenario was initially analyzed by Yu {\it et al.} in Ref.~\cite{yu_spatio-temporal_1995}. Figure~\ref{fig1} shows the modification of the STSs all along $\mathcal{B}_a$ [see labels (i)-(iii)] and $\mathcal{B}_b$ [see labels (iv)-(vi)], respectively. Figures~\ref{fig2}(i).1-(iii).1 show the reconstruction of the $\mathcal{B}_a$-related STSs using the solution ansatz (\ref{ansatz1}) for $E=4,6$ and $8$, where the values of $a$, $\eta$ and $I_{peak}$, obtained from Eq.~(\ref{Eq_for_a}) and Eq.~(\ref{eq2_static}), are shown in Table~\ref{table1}.  To represent the STS we plot isosurfaces for different intensity values (see caption in Fig.~\ref{fig2}). Figures~\ref{fig2}(i).2-(iii).2 represent the wave-function intensity cross-sections at the plane $t=0$, i.e. $I_\sigma\equiv I(x,y,t=0)$. Increasing $E$, from $E_0$ to $E_f$, the STSs on $\mathcal{B}_a$ decrease their spatial width $a$ and temporal width $\eta^{-1}$ [see how $\eta$ increases in Fig.~\ref{fig1}(b)],  while increasing their amplitude, i.e., their peak intensity.  
After crossing $E_f$, the $\mathcal{B}_b$-STSs just continue to decrease in $a$ and $\eta^{-1}$, while increases drastically in $I_{peak}$ [see Figs.~\ref{fig2}(iv)-(vi)]. As a result, the $\mathcal{B}_b$-related STS compresses in all three dimensions, becoming a singularity with decreasing $E$. This process can be appreciated in Figs.~\ref{fig2}(iv).2-(vi).2.

\subsection{Guiding, self-defocusing material with normal GVD}
In the case of a guiding, self-defocusing nonlinear material ($\nu=-1$) with normal GVD ($\delta=-1$), Eqs.~(\ref{eq1_static}), 
(\ref{eq2_static}), and (\ref{Eq_inten})
become
\begin{eqnarray}
E=2\pi a\sqrt{-6(1+\rho a^4)}, 
\end{eqnarray}
\begin{eqnarray}
\eta=\frac{1}{2 a}\sqrt{-6(1+\rho a^4)}, 
\end{eqnarray}
\begin{eqnarray}
I_{peak}=-\frac{3}{a^2}(1+\rho a^4).
\end{eqnarray}

Figures~\ref{fig1}(d)-(f) show the modification of these quantities as a function of $E$ for $\rho=-1$. In this regime, Eq.~(\ref{Eq_for_a}) has a single real solution, and the solution ansatz parameters are single-valued in $E$. Thus, here there exists just a single STS for any fixed value of $E$.
This scenario was partially analyzed by Raghavan and Agrawal in \cite{raghavan_spatiotemporal_2000}. In contrast with the anomalous GVD/self-focusing case, we can see that for the same energy interval, the STS is spatially wider than the CW state as shown, in Fig.~\ref{fig1}(d). Furthermore, when compared with its anomalous GVD analog, this STS is wider in space, thinner in time, and possesses a lower peak intensity.

\section{Effective dynamics of spatiotemporal in the Lagrangian formulation}\label{sec:4}
So far, we have just described the properties of steady-state STSs, but we do not know anything yet about their dynamics while propagating in $z$, nor their stability. In order to do that, we need to include the dependence of $z$ on the STS description. Here we derive a four-dimensional (4D) effective dynamical system in the independent evolution variable $z$, by considering the Lagrangian formalism. To do so, first, we have to generalize the solution ansatz to include the contribution of $z$. 

Our new ansatz is a product of the static one [see Eq.~(\ref{ansatz1})] and a spacetime-dependent phase contribution, namely
\begin{equation}\label{ansatz_chirp}
u(z,x,y,t)\equiv v[x,y,t;q_A(z)] {\rm Exp}\left(iC[x,y,t;q_B(z)]\right),
\end{equation}
with $q_A(z)=\{\eta(z),a(z)\}$, $q_B(z)=\{\theta(z),\alpha(z),\phi(z)\}$, and
$$C[x,y,t;q_B(z)]\equiv t^2 \theta (z)+\left(x^2+y^2\right) \alpha (z)+\phi (z),$$
where $\alpha$ represents the spatial chirp, $\theta$ the temporal chirp, and $\phi$ the phase. 

This ansatz leads to the effective $z$-dependent Lagrangian
\begin{equation}
\begin{aligned} 
	-L(z)E^{-1}=&d_z\phi+\frac{\pi^2d_z\theta}{12\eta^2}+a^2d_z\alpha+\frac{\delta}{6}\eta^2+\frac{\pi^2\delta\theta^2}{6\eta^2}\\
	+&(4\alpha^2-\rho)\frac{a^2}{2} 
	+ \frac{1}{2a^2}\left(1-\frac{\nu E\eta}{6\pi}\right),
\end{aligned}
\label{red_Lag}
\end{equation}
where $q(z)=\{q_A(z),q_B(z)\}=\{\eta(z),a(z),\theta(z),\alpha(z),\phi(z)\}$.
The Euler-Lagrange equations associated with $\theta,\alpha,\eta,$ and $a$
lead to the 4D effective dynamical system 
\begin{equation}\label{DS}
\begin{array}{l}
	\displaystyle\frac{d\eta}{dz}=f_1\equiv-2\delta\eta\theta,\\\\
	\displaystyle\frac{da}{dz}=f_2\equiv2 a \alpha,\\\\	
	\displaystyle\frac{d\theta}{dz}=f_3\equiv2\delta\left(\frac{\eta^4}{\pi^2}-\theta^2\right)-\frac{E\nu\eta^3}{2\pi^3 a^2},\\\\
	\displaystyle\frac{d\alpha}{dz}=f_4\equiv\frac{1}{2 a^4}\left(1-\frac{\nu\eta E}{6\pi}\right)+\frac{\rho}{2}-2\alpha^2,
\end{array}
\end{equation}
which can be also written in a more compact fashion as 
\begin{equation}
\frac{d q}{d z}=f(z;E), 
\end{equation}
where $f=(f_1,f_2,f_3,f_4)$ is the nonlinear vector field defined in (\ref{DS}). Note that Eq.~(\ref{red_Lag}) depends on $\phi$ only through $d_z\phi$ and therefore all the terms in the associated Euler-Lagrange equation are null. 

With this approach, we have been able to reduce the infinite-dimensional Eq.~(\ref{GPE}) down to a 4D dynamical system, which describes the evolution of the four STS parameters along the propagation distance $z$.
The Euler-Langrange equation associated with $\phi$ does not give any useful information, since each of its components is null. Therefore, the contribution of the phase remains irrelevant.  

Here the fixed points, or equilibria of the system, $q=q_e\equiv(\eta_e,a_e,\alpha_e,\theta_e)$ satisfy $dq_e/dz=0$. This condition leads to two types of equilibria, which are both chirp-free (i.e., $\theta_e=\alpha_e=0$). The simplest one corresponds to the CW beam, which is homogeneous in time. This equilibrium reads $q_{e}^h=(\eta^h_e,a^h_e,\alpha^h_e,\theta^h_e)\equiv(0,1,0,0)$.
The CW solution is plotted, for both regimes, in Fig.~\ref{fig1} by using a solid blue line.
The other equilibrium solution is localized in spacetime, and it corresponds to STSs $q=q_e^s=(\eta_e^s,a_e^s,\alpha_e^s,\theta_e^s)$, where $(\alpha_e^s,\theta_e^s)=(0,0)$, and $a_e^s$ and $\eta_e^s$ satisfy Eqs.~(\ref{eq1_static}) and (\ref{eq2_static}), respectively. 

In Section~\ref{sec:6.3} we will perform a linear stability analysis of Eq.~(\ref{DS}) around its equilibria, and study how the linear dynamics of the system change as a function of $E$. These findings will be later confirmed by numerically solving Eq.~(\ref{DS}).

\section{Effective dynamics of spatiotemporal solitons in the Hamiltonian formulation}\label{sec:5}
Most of the time, when seeking STS solutions of nonlinear partial differential equations, the variational Ritz optimization method has focused on a Lagrangian description \cite{yu_spatio-temporal_1995,raghavan_spatiotemporal_2000}. In this section, we show that the effective 4D   
dynamical system (\ref{DS}) and their equilibria can also be obtained in the framework of the Hamiltonian formalism. 

For a start, we need to introduce the generalized momenta $p=(p_\eta,p_a,p_\theta,p_\alpha,p_\phi)$, which are canonically conjugate of $q=(\eta,a,\theta,\alpha,\phi)$. These momenta are defined as 
\begin{eqnarray*}
p_\eta=\frac{\partial L}{\partial (d_z\eta)}, \qquad  p_a=\frac{\partial L}{\partial (d_za)}, \qquad  p_\theta=\frac{\partial L}{\partial (d_z\theta)}, 
\end{eqnarray*}
\begin{eqnarray*}
p_\alpha=\frac{\partial L}{\partial (d_z\alpha)}, \qquad p_\phi=\frac{\partial L}{\partial (d_z\phi)}.
\end{eqnarray*}
By utilizing the Lagrangian function defined by Eq.~(\ref{red_Lag}), the momenta become $p_\eta=0$, $p_a=0$ and
\begin{equation}\label{momenta}
p_\theta=-\frac{E\pi^2}{12\eta^2},\qquad   p_\alpha=-Ea^2,\qquad  p_\phi=-E.
\end{equation}

At this point, the Hamiltonian can be computed by means of two different approaches. One of them uses Eq.~(\ref{efec_hamilton}) with the Hamiltonian density (\ref{Hamil_density}) and the chirp-dependent ansatz (\ref{ansatz_chirp}). The other option consists of applying the Legendre transform to Eq.~(\ref{red_Lag}) and using the generalized momenta in Eq.~(\ref{momenta}). In any case, we obtain the effective Hamiltonian as
\begin{eqnarray*}
H=E  \left[\frac{\delta\eta^2}{6}+\frac{\pi^2\delta\theta^2}{6\eta^2}+\frac{a^2}{2}(4\alpha^2-\rho)+\frac{1}{2a^2}\left(1-\frac{E\nu\eta}{6\pi}\right)\right],  
\end{eqnarray*}
which, when written in terms of generalized momenta, reads as
\begin{eqnarray*}
-H(\theta,\alpha,p_\theta,p_\alpha)E^{-1}=\frac{\delta E \pi^2}{72p_\theta}+\frac{2\delta\theta^2p_\theta}{E}+\frac{p_\alpha}{2E}(4\alpha^2-\rho)\\+\frac{E}{2p_\alpha}\left(1-\frac{E\nu}{6}\sqrt{-\frac{E}{12p_\theta}}\right)
\label{eq_gen_memta_H}
\end{eqnarray*}

In this case, the Hamiltonian equations of motion describing the dynamics of the system are 
\begin{equation}\label{DS2}
\begin{array}{l}
	\displaystyle\frac{dp_\theta}{dz}=-\frac{\partial H}{\partial\theta}=4\delta\theta p_\theta,\\\\
	\displaystyle\frac{dp_\alpha}{dz}=-\frac{\partial H}{\partial\alpha}=4\alpha p_\alpha,\\\\	
	\displaystyle\frac{d\theta}{dz}=\frac{\partial H}{\partial p_\theta}=-2\delta\theta^2+\frac{E^2}{72p_\theta^2}\left(\delta\pi^2+\frac{E^2\nu}{4p_\alpha}\sqrt{-\frac{12 p_\theta}{E}}\right),\\\\
	\displaystyle\frac{d\alpha}{dz}=\frac{\partial H}{\partial p_\alpha}=\frac{1}{2}(\rho-4\alpha^2)+\frac{E^2}{2p_\alpha^2}\left(1-\frac{E\nu}{6\pi}\sqrt{-\frac{E\pi^2}{12 p_\theta}}\right),
\end{array}
\end{equation} 
This dynamical system, defined in the phase space $(p_\theta,p_\alpha,\theta, \alpha)$, possesses the same information as Eq.~(\ref{DS}), which describes the dynamics of the system in the $(\eta,a,\theta,\alpha)$ space. Moreover, the equation $\frac{d p_\phi}{dz}=-\frac{\partial H}{\partial \phi}=0$, implies that $p_\phi=-E$ remains constant during propagation, which means that $E$ is conserved in the course of the $z$-evolution.

In this formulation, the equilibria of the system are obtained from the nullity of the gradient of $H$, evaluated at $(q,p)=(q_e,p^e)$ \cite{abraham_foundations_2008}, namely
\begin{equation}
\mathcal{D}H|_{q_e}\equiv(\partial_\theta H,\partial_\alpha H,\partial_{p_\theta} H,\partial_{p_\alpha} H)_{(q_e,p^e)}=0.
\end{equation}
The first two conditions yield $\theta_e=\alpha_e=0$. These, once combined with  
the steady-state versions of the third and fourth equations in (\ref{DS2}), lead to
\begin{equation}
\delta\pi^2+\frac{E^2\nu}{4p^e_\alpha}\sqrt{-\frac{12 p^e_\theta}{E}}=0,
\end{equation}
\begin{equation}
\frac{\rho}{2}+\frac{E^2}{2(p^e_\alpha)^2}\left(1-\frac{E\nu}{6\pi}\sqrt{-\frac{E\pi^2}{12 p^e_\theta}}\right)=0,
\end{equation}
respectively. By inserting the expressions for $p_\alpha$ and $p_\theta$ in the previous equations, one recovers the STS solution conditions (\ref{fixed_1}) and (\ref{fixed_2}). In Section~\ref{sec:6.2} we will use this formulation to estimate the stability of STSs by using the Lyapunov stability criterion. This approach was also considered in \cite{shtyrina_coexistence_2018}, but using a simpler Gaussian ansatz.

\section{Spatiotemporal solitons stabiltiy}\label{sec:6}
So far, we have studied the shape, features, and existence regions of STSs, without mentioning their stability properties: now, the time for tackling this issue has arrived. Different complementary approaches exist in order to determine STS stability, although most authors decide to use just one of them and neglect the information which could be gained from other methods. In this section, we determine the stability of STSs by using three different methods, which are of common use. They are the Vakhitov-Kolokolov stability criterion \cite{vakhitov_stationary_1973}, the Lyapunov stability criterion \cite{reiszig_j_1962}, and the spectral stability criterion \cite{wiggins_introduction_2003}.

\begin{eqnarray*}
	H_e\equiv H(\theta_e,\alpha_e,p^e_\theta,p^e_\alpha)=\qquad\qquad\qquad\qquad\qquad\qquad\\-E\left[\frac{\delta E \pi^2}{72p^e_\theta}-\frac{p^e_\alpha}{2E}\rho+\frac{E}{2p^e_\alpha}\left(1-\frac{E\nu}{6}\sqrt{-\frac{E}{12p^e_\theta}}\right)\right],
\end{eqnarray*}
or, in terms of the generalized coordinates, as
\begin{figure}[!t]
	\centering	\includegraphics[width=\columnwidth]
	{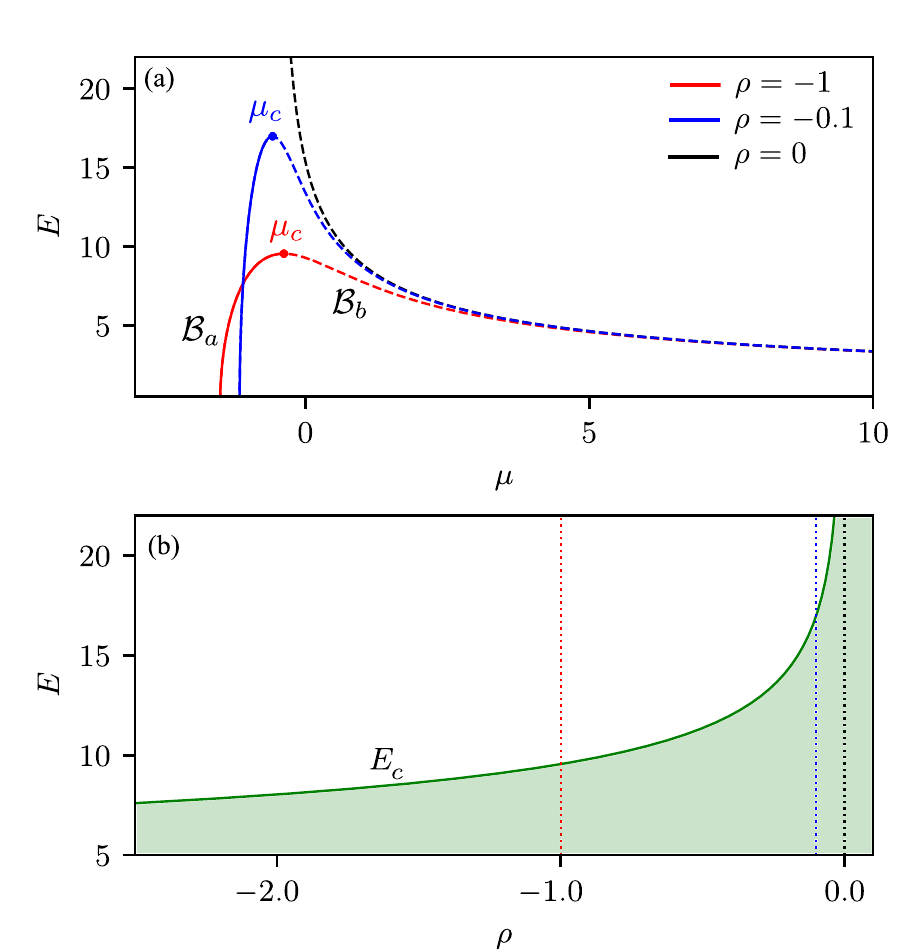}
	\caption{(a) Dependence of the energy $E$ with $\mu$ for different values of $\rho$ in the anomalous GVD regime. Stable (unstable) branches are plotted by using solid (dashed) lines. (b) Region of existence of STSs as a function of $\rho$ (see green shadowed area). The line limiting that area is $E=E_c$. Vertical dashed lines correspond to the three cases plotted in (a). }
	\label{fig_mu}       
\end{figure}
\subsection{Vakhitov-Kolokolov stability criterion}\label{sec:6.1}
The dependence of the propagation constant $\mu$ on energy $E$ allows us to determine the linear stability of STSs, in terms of the Vakhitov-Kolokolov stability (VKS) criterion \cite{vakhitov_stationary_1973,kivshar_optical_2003}. Such a dependence was computed in Secs.~\ref{sec:3}, and it is graphically illustrated in Fig.~\ref{fig_mu}(a) for the anomalous dispersion propagation regime. The VKS criterion establishes that an STS state is linearly stable (i.e., with respect to small perturbations), if the derivative of the energy with respect to $\mu$ is a positive quantity (i.e., $dE/d\mu>0$), and unstable otherwise. 
The main idea behind this criterion is based on the analysis of the properties of the linear operator associated with Eq.~(\ref{GPE}), evaluated on the soliton solution: we recommend the interested reader to consult at Secs.~2.3 in  \cite{kivshar_optical_2003} for details. 

The red curve in Fig.~\ref{fig_mu}(a) correspond to $\rho=-1$, the same case that we have studied in previous sections [see  Figs.~\ref{fig1}(a)-(c)]. This criterion shows that $\mathcal{B}_a$ is stable (see solid line), while $\mathcal{B}_b$ is unstable (see dashed line).
The instability threshold $\mu_c$ occurs whenever $dE/d\mu|_{\mu_c}=0$, which leads to 
\begin{equation}
\mu_c=-\frac{\sqrt{-3\rho}}{2}+\frac{2}{3}\left(\sqrt{-3\rho}-1\right), \qquad E_c=\frac{4\pi}{(-3\rho)^{1/4}}.
\end{equation}
This point corresponds to the fold occurring at $E_f$: in the following, we shall write $(\mu_c,E_c)=(\mu_f,E_f)$. This point is marked by means of a red bullet in Fig.~\ref{fig_mu}(a).  

So far, we have studied the formation of STSs whenever $\rho=-1$. However, it remains to be studied how the stability and the region of existence of STSs are modified for different values of $\rho$. To unveil these changes, we show, in Fig.~\ref{fig_mu}(a), the modification of the $\mu$ curve for other two characteristic values of $\rho$: specifically, $\rho=-0.1$ (in blue) and $\rho=0$ (in black). By increasing $\rho$, the region of existence of STSs broadens, as the critical point $(E,\mu)=(E_c,\mu_c)$ moves towards higher values of $E$. 
The dependence of $E_c$ upon $\rho$ is shown in Fig.~\ref{fig_mu}(b). Here, the green shadowed area corresponds to the region of existence of STSs. 
For $\rho=0$, only the branch $\mathcal{B}_b$ survives, and the STSs are always unstable. 
These results show that the presence of a non-vanishing parabolic potential is essential for the stabilization of STSs \cite{shtyrina_coexistence_2018}.  
A similar analysis may show that, in the normal GVD regime [see  Figs.~\ref{fig1}(d)-(f)], the single branch of STS solutions remains always stable. 

\subsection{Lyapunov stability criterion}\label{sec:6.2}
The Hamiltonian provides information about the stability of the fixed points in terms of the Lyapunov stability criterion \cite{reiszig_j_1962,rosen_particlelike_1965}. 
Whenever it is evaluated at 
the STS equilibria, 
the Hamiltonian reads as 

\begin{equation}
H_e\equiv H(q_e)=E\left[\frac{\delta\eta_e^2}{6}-\frac{a_e^2}{2}\rho+\frac{1}{2a^2}\left(1-\frac{E\nu\eta_e}{6\pi}\right)\right].  
\end{equation}
Next, the Lyapunov stability criterion establishes that, if an equilibrium $q_e$ minimizes (maximizes) $H$, such a state is stable (unstable). The way of determining if $q_e$ maximizes or minimizes $H$, is by studying the determinant of the Hessian matrix associated with $H$, once that it is evaluated at such a point.
Defined in terms of its components, the Hessian matrix of $H$ evaluated at $q_e$, reads as
\begin{equation}
\mathcal{D}^2H(q_e)_{(i,j)}\equiv\left(\frac{\partial^2 H}{\partial q_i\partial q_j}\right)(q_e),
\end{equation} 
where the subindex $i,j=1,\cdots,4$, scan the four STS parameters $(q_1,q_2,q_3,q_4)=(\eta,a,\theta,\alpha)$.
\begin{figure}[!t]
\centering
\includegraphics[scale=1.]{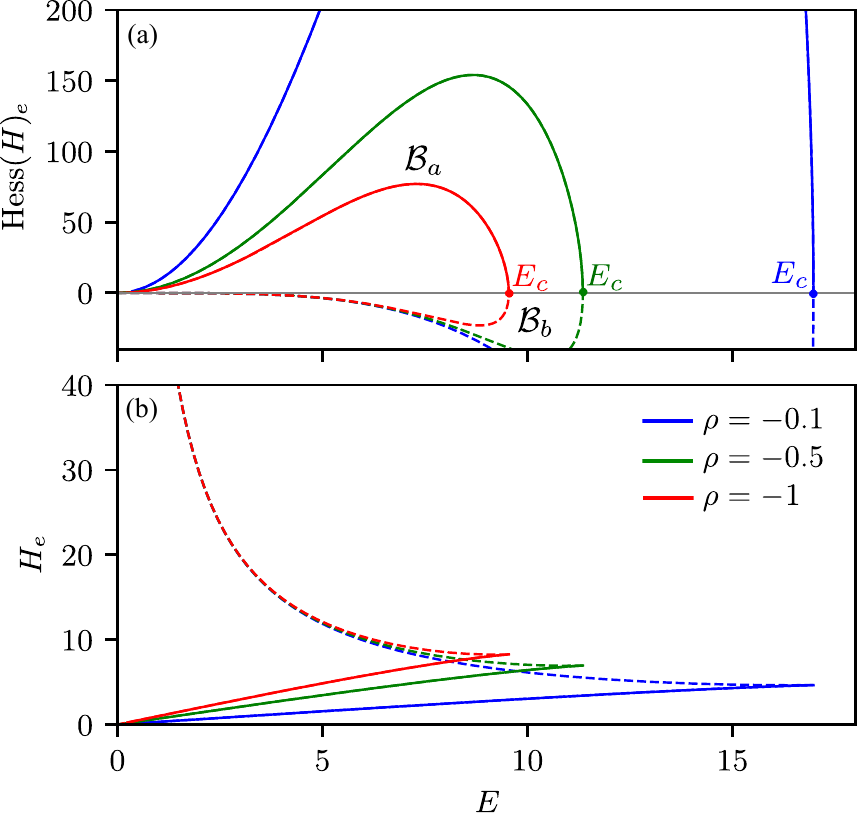}
\caption{Lyapunov stability of STSs. Panel (a) shows the dependence of the Hessian of $H$ upon $E$, for the self-focusing/anomalous GVD regime, and three values of $\rho$. Panel (b) shows the $H$ versus $E$ diagram. }
\label{fig3}    
\end{figure}
The determinant of this matrix, known as the Hessian of $H$, reduces to
\begin{eqnarray*}
{\rm Hess}(H)_e\equiv {\rm det}\left(\mathcal{D}^2H(q_e)\right)=\qquad\qquad\qquad\qquad\qquad\\-\frac{\delta  E ^4 \left(12 \pi ^2 a_e^6 \delta  \rho +6 \pi  a_e^2 \delta  (\eta_e \nu  E -6 \pi )+ E ^2\right)}{27 a_e^4 \eta_e^2}. 
\end{eqnarray*}
The main thing that we need to know now is that, if ${\rm Hess}(H)_e>0$, $H_e=H(q_e)$ is a minimum of $H$, and $q=q_e$ is a stable equilibrium. However, when ${\rm Hess}(H)_e<0$, $H$ has a maximum at $q=q_e$, which corresponds to an unstable STS. The transition between these two situations occurs when 
${\rm Hess}(H)_e=0$, a condition which defines the instability threshold.

Figure~\ref{fig3}(a) shows ${\rm Hess}(H)_e$ as a function of $E$, for the case of anomalous GVD/self-focusing regime and $\rho=-1$ (see red curve). The solid portion of this curve (i.e., ${\rm Hess}(H)_e>0$) corresponds to the stable STS branch $\mathcal{B}_a$, which extends from $a=a_{e}^h$ up to $a=a_f$ [see Fig.~\ref{fig1}(a)]. Whereas the dashed section (${\rm Hess}(H)_e<0$) corresponds to  $\mathcal{B}_b$. In this case, the condition ${\rm Hess}(H)_e=0$ corresponds to the turning point or fold of the STS solutions, which occurs at $E=E_c=E_f$. Thus, the prediction of the Lyapunov stability criterion agrees with the previously described VKS criterion (see Section~\ref{sec:6.1}). For a comparison, in Figure~\ref{fig3}(a) we also trace ${\rm Hess}(H)_e$ for other values of $\rho$. Figure~\ref{fig3}(b) illustrates how $H_e$ changes with $E$, for several values of $\rho$. For a given value of $E$, the minimum attained by $H_e$ corresponds to a stable solution on $\mathcal{B}_a$. Whereas the maximum $H_e$ corresponds to an unstable state on $\mathcal{B}_b$, as predicted by the Hessian of $H$. The cusp of this graph corresponds to the position of the fold point, which is shown in Figs.~\ref{fig1}(a)-(d). These graphs are known as $H$ vs. $E$ diagrams and are a fast method to determine stability \cite{akhmediev_hamiltonian-versus-energy_1999}.


In the normal regime, ${\rm Hess}(H)_e>0$ for every value of $E$: thus, STSs are always stable.

\subsection{Spectral linear stability and types of equilibria}\label{sec:6.3}

The spectral linear stability analysis is based on the computation of the set of eigenvalues (i.e., the spectrum) associated with the linearization of Eq.~(\ref{DS}) around the fixed point $q_e$. This analysis allows us to determine how the equilibria of the system react against perturbations of the form $q=q_e+\epsilon \tilde{q}$, where $\epsilon\ll 1$ and  $\tilde{q}\equiv(\tilde{\eta},\tilde{a},\tilde{\theta},\tilde{\alpha})$. Moreover, one may classify them according to their behavior.  

Very close to a fixed point $q_e$, the dynamics of the system (\ref{DS}) are captured by the linear dynamical system
\begin{eqnarray}
\frac{d \tilde{q}}{dz}=\mathcal{J}[q_e]\tilde{q},
\end{eqnarray}
where $\mathcal{J}[q_e]$ is the Jacobian matrix of the vector field $f$ [see Eq.~(\ref{DS})], which is defined by its components as follows 
\begin{eqnarray}
\mathcal{J}[q_e]_{(i,j)}\equiv\mathcal{D}f_{(i,j)}(q_e)=\left(\frac{\partial f_i}{\partial {q_i}}\right)(q_e).
\end{eqnarray}

In our case, the Jacobian matrix becomes
\begin{equation}\label{Jacob}
\mathcal{J}[q_e]\equiv\left[\begin{array}{cccc}
	-2\delta\theta_e &0&-2\delta\eta_e&0 \\ 
	0&2\alpha_e&0&2 a_e\\
	\mathcal{J}_{31} &\frac{E\nu\eta_e^3}{\pi^3a_e^3} &-4\delta\theta_e&0 \\
	-\frac{\nu E}{12\pi a_e^4}&\mathcal{J}_{42}&0&-4\alpha_e
\end{array}\right],
\end{equation}
with
\begin{eqnarray*}
\mathcal{J}_{31}\equiv\left(8\delta\eta_e-\frac{3}{2}\frac{E\nu}{\pi a_e^2}\right)\frac{\eta_e^2}{\pi^2}, \qquad   \mathcal{J}_{42}\equiv-\frac{2}{a_e^5}\left(1-\frac{\nu E\eta_e}{6\pi}\right).
\end{eqnarray*}

Then, the stability of the fixed points can be evaluated by solving the linear eigenvalue problem \begin{equation}
\mathcal{J}w=\lambda w,
\end{equation}
where $\lambda$ and $w$ are the eigenvalue and eigenvector associated with the Jacobian (\ref{Jacob}), respectively. 
The eigenvalues satisfy the bi-quadratic characteristic polynomial
\begin{equation} \lambda^4+c_2\lambda^2+c_0=0,
\end{equation}
with the coefficients
\begin{eqnarray*}
c_2\equiv 2(\delta_e\eta_e\mathcal{J}_{31}-a_e\mathcal{J}_{42}), 
\end{eqnarray*}
\begin{eqnarray*}
c_0\equiv -\delta\left(4a_e\eta_e\mathcal{J}_{31}\mathcal{J}_{42}+\frac{E^2\eta^4}{3\pi^4 a_e^6}\right),
\end{eqnarray*}
\begin{figure}[!t]
\centering	\includegraphics[scale=0.9]
{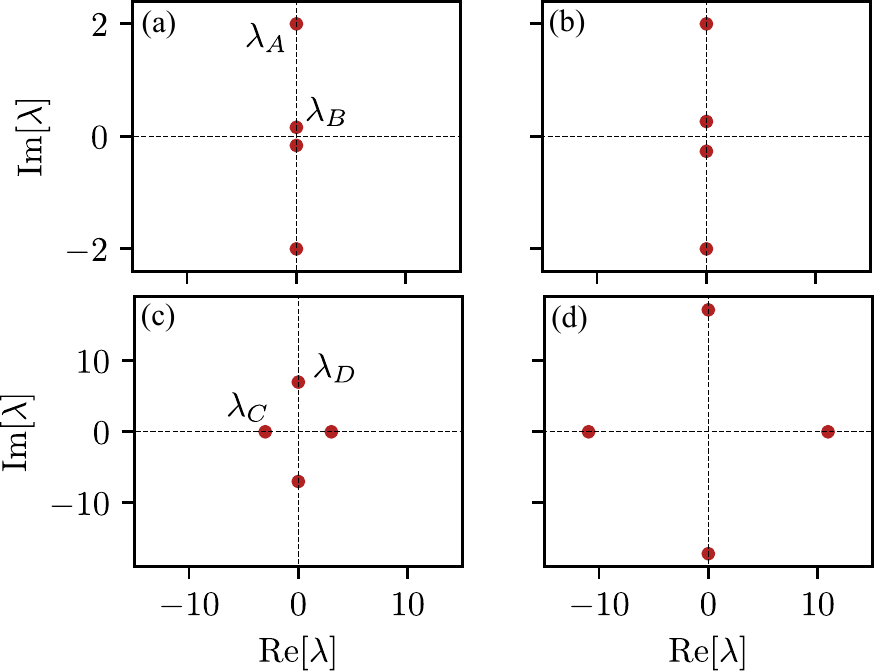}
\caption{Eigenvalues associated with the dynamical system (\ref{DS}) for the anomalous GVD regime. Panels (a) and (b) show the spectrum associated with the center equilibria in $\mathcal{B}_a$ for $E=4$ and $E=6$ in Fig.~\ref{fig1}(d), respectively. Panels (c) and (d) show the eigenvalues associated with the saddle-center points on $\mathcal{B}_b$ for the same energy values.  }
\label{Eigen_anomal}       
\end{figure}

which can be easily solved to obtain
\begin{equation}
\lambda=\pm\sqrt{\frac{-c_2\pm\sqrt{c_2^2-4c_0}}{2}}.
\end{equation}
In the following, we will study the linear stability of the equilibria, for the two regimes under study.

\begin{figure*}[!t]
\centering	\includegraphics[scale=1]{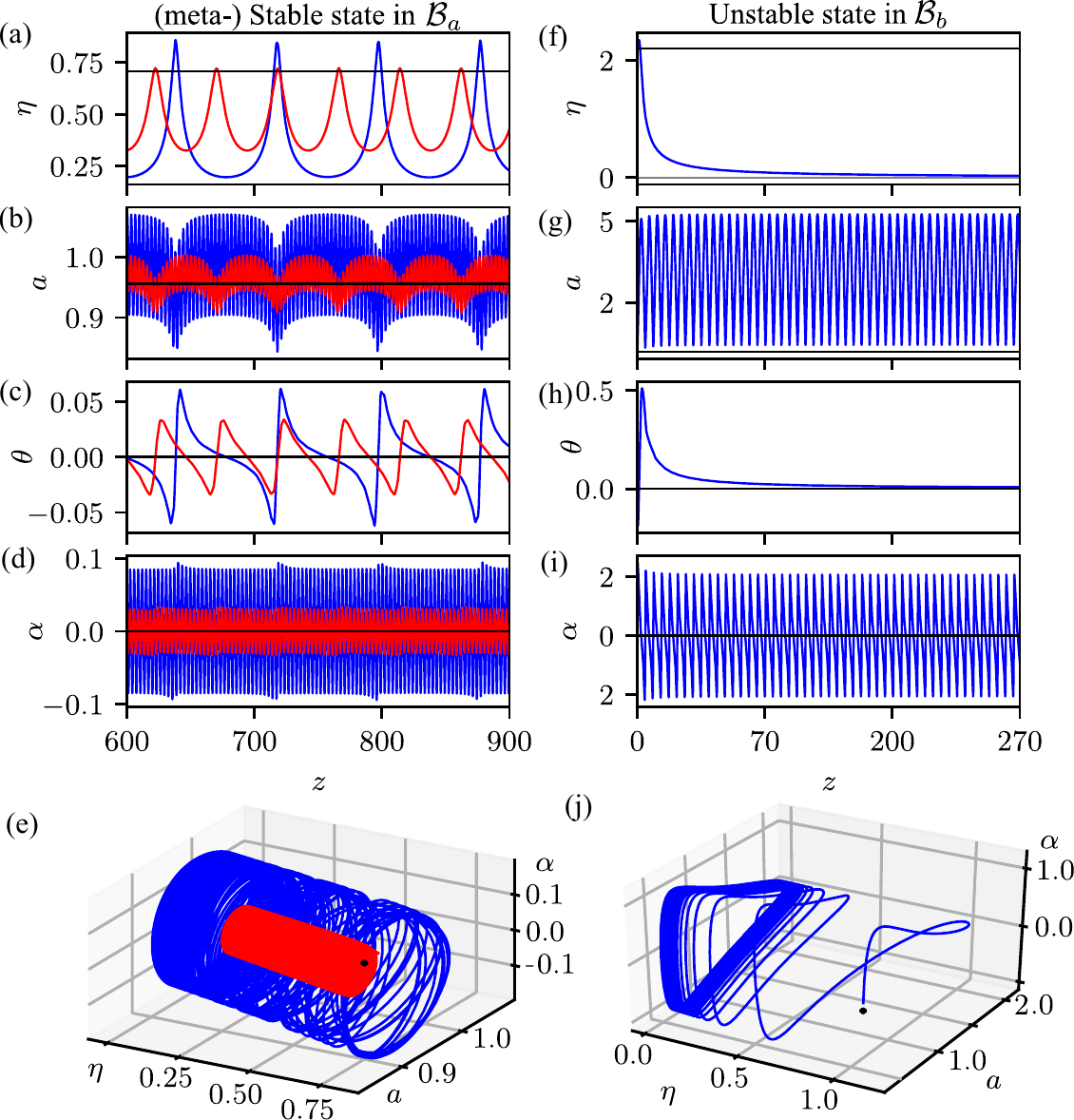}
\caption{Panels (a)-(d) illustrate the system dynamics, when considering as the initial condition a metastable equilibrium on the $\mathcal{B}_a$-STS branch for $E=6$, slightly perturbed by a constant term in all the parameter directions. The red curve corresponds to $\epsilon=0.001$, and the blue curve to $\epsilon=0.01$. The black horizontal line shows the analytically predicted equilibrium $q_e=(\eta_e,a_e,\theta_e,\alpha_e)$. In panel (e) we project these dynamics on the 3D subspace $\{(\eta,a,\alpha)\}$. Panels (f)-(i) shows the $z$-evolution of an unstable equilibrium on $\mathcal{B}_b$ for $E=6$. Panel (j) shows its 3D projection. }
\label{fig5}       
\end{figure*}
Figures~\ref{Eigen_anomal}(a) and \ref{Eigen_anomal}(b) show the eigenspectrum associated with stable STSs on $\mathcal{B}_a$ for $E=4$ and $E=6$, respectively [see Fig.~\ref{fig2}(i)-(ii)]. The spectrum consists of four pure imaginary eigenvalues $\sigma=\{\pm i\lambda_A,\pm i\lambda_B\}$, with $\lambda_A>\lambda_B>0$. 
A fixed point with these eigenvalues is known as a {\it center} \cite{wiggins_introduction_2003}. Center points are neutrally stable, in the sense that nearby trajectories (i.e., soliton parameter perturbations) are neither repelled nor attracted to it, but they undergo permanent oscillations.


In contrast, eigenspectra for STSs on $\mathcal{B}_b$ are shown in Figs.~\ref{Eigen_anomal}(c) and \ref{Eigen_anomal}(d): they correspond to the STS which are plotted in Figs.~\ref{fig2}(v) and \ref{fig2}(vi) for $E=4$ and $E=6$, respectively. These spectra are formed by two pure imaginary and two pure real eigenvalues, namely, with $\sigma=\{\pm\lambda_C,\pm i\lambda_D\}$. In this case, the associated equilibria are known as {\it saddle-centers}, and are unstable \cite{wiggins_introduction_2003}. Let us check these results by performing numerical simulations of the evolution of the dynamical system (\ref{DS}). The outcome of these simulations is shown in Fig.~\ref{fig5}.

Figures~\ref{fig5}(a)-(d) show, by using a black line, the constant $z$-evolution of a center unperturbed state on $\mathcal{B}_a$ for $E=6$. The $z$-evolution of a constant perturbation $\tilde{q}$ on the STS (i.e., $q=q_e+\epsilon\tilde{q}$) is shown in the same figure for $\epsilon=0.01,0.1$ by using a red and a blue line, respectively. For any of these perturbations, the center metastable equilibrium evolves towards  
periodic oscillations with two different frequencies, where $\eta$ and $\theta$ oscillate with a lower frequency [see Figs.~\ref{fig5}(a),(b)], whereas $a$ and $\alpha$ oscillate with a higher frequency [see Figs.~\ref{fig5}(c),(d)], as predicted by the linear theory.
By increasing $\epsilon$, so does the amplitude of the oscillations in all of the $q_e$ components, while the oscillation frequency decreases. Note that with increasing $\epsilon$, the variable $\theta$ develops relaxation oscillations [see Fig.~\ref{fig5}(c)]. 

In the phase space, these oscillations correspond to the closed orbit which is illustrated through a 3D projection on the subspace spanned by $\{(\eta,a,\theta)\}$ in Figs.~\ref{fig5}(e).
In contrast to the limit cycles (which are typical for dissipative systems), these orbits are not isolated, but form a continuous family around the center equilibrium \cite{raghavan_spatiotemporal_2000}. In this sense, different perturbations lead to different oscillatory STSs, which coexist for the same value of energy.
We find that by increasing $E$, the STS center and the periodic oscillations modify in shape, amplitude, and periodicity. 

\begin{figure*}[tbp]
\centering
\includegraphics[scale=1]{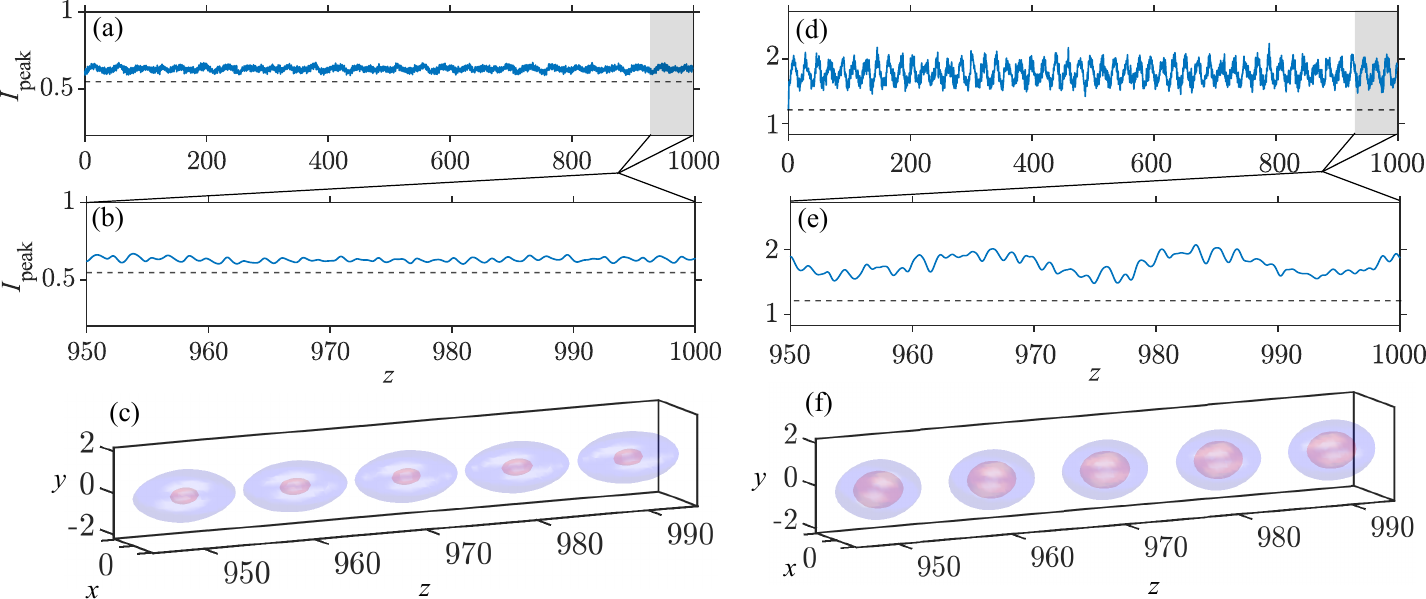}		
\caption{Evolution with distance $z$ of a stable STS for $E=6$ [see (a), (b), (c)] or $E=8$ [see (d), (e), (f)]. Panel (a) shows the variation of the peak STS intensity vs. the propagation distance. Panel (b) shows a close-up view of (a) for the interval $z\in[950,1000]$. Panel (c) shows the evolution of the STS along the interval shown in (b), obtained by plotting two isosurfaces at $I_1=0.5$ (red), and $I_2=0.1$ (blue). The dashed gray straight line in (a) and (b) represents the theoretical value of the STS intensity. Panels (d), (e), (f) show the same information as (a), (b), (c), but for $E=8$. }
\label{3D_simu_E6}       
\end{figure*}

Figures~\ref{fig5}(f)-(i) show the evolution of the system (\ref{DS}), when taking as the initial condition an unstable saddle-center STS fixed point with $E=6$. Here the parameters describing the temporal part  of the STS (i.e., $\eta$ and $\theta$) evolve rapidly to $(\eta,\theta)=(0,0)$ as they follow the eigenvectors associated with $\pm\lambda_C$, while $a$ and $\alpha$ undergo periodic oscillations following the dynamics dictated by the pure imaginary eigenvalues $\pm i\lambda_D$. The 3D projection of this dynamics on the subspace $\{(\eta,a,\theta)\}$ yields the trajectory shown in Fig.~\ref{fig5}(j), where the black dot is the initial condition, corresponding to the unstable saddle-center point. This representation clearly shows that the system evolves to an oscillatory state in the plane $\eta=0$, which corresponds to the CW state. Thus, we may interpret this behavior as beam self-imaging \cite{karlsson_dynamics_1992}.

In the normal GVD/self-defocusing scenario, the eigenspectrum is of the form $\sigma=\{\pm i\lambda_A,\pm i\lambda_B\}$, as it occurs for the $\mathcal{B}_a$ solution branch, and the dynamics around such equilibria is similar to what we have depicted in Figs.~\ref{fig5}(a)-(e).

\section{Full three-dimensional numerical simulations}\label{sec:7}
The aim of this section is to compare the previously discussed theoretical results with direct numerical solutions of the original GPE [see Eq.~(\ref{GPE})]. To solve this initial value problem, we take as the initial condition the approximate variational solution defined by Eq.~(\ref{ansatz_chirp}) with the parameters corresponding to equilibria of the effective dynamical theory. To do so, we utilize a pseudo-spectral split-step algorithm \cite{agrawal_applications_2008}, where the differential part of Eq.~(\ref{GPE}) is evaluated via a fast Fourier transform and the linear potential and the nonlinear term are computed exactly as a phase shift. 
Moreover, we confirm the validity of our results by implementing Runge-Kutta and predictor/corrector simulation schemes \cite{frolkovic_numerical_1990}. 
In what follows, we analyze each scenario separately.

\subsection{Anomalous/self-focusing scenario}\label{sec:7.2}
\begin{figure}[!t]
\centering	\includegraphics[width=\columnwidth]{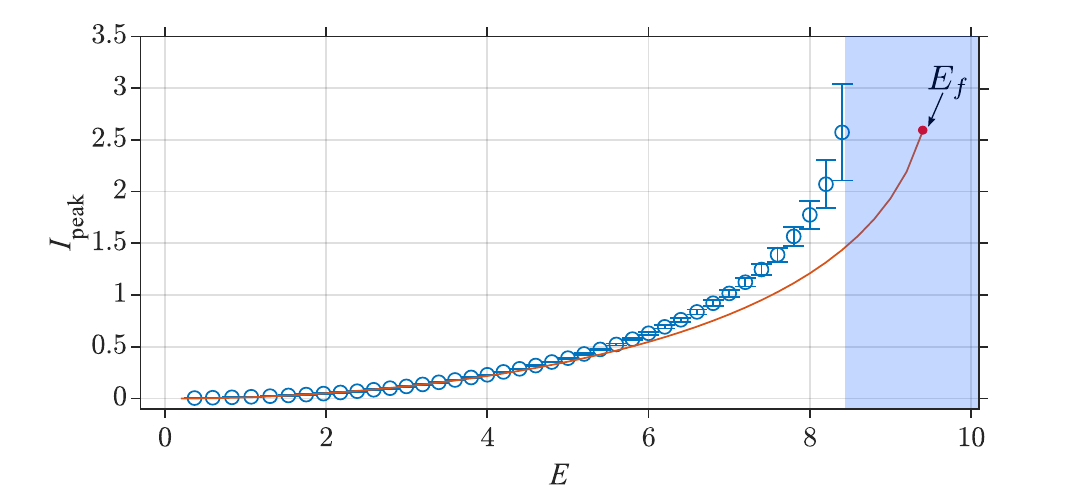}
\caption{Evolution of peak intensity of stable STS with the energy $E$ in the anomalous GVD/self-focusing regime (i.e., $\delta=\nu=1$). The red line shows the analytical value, while the blue circles and the error bars represent the average intensity value and the standard deviation for stable states, respectively, which are obtained from full 3D numerical simulations.
}
\label{fig_peak_vs_e}      
\end{figure}

The $z$-evolution of the initial stable chirp-free STS solution is shown in Figs.~\ref{3D_simu_E6}(a)-(c) and \ref{3D_simu_E6}(d)-(f) for $E=6$ and $E=8$, respectively. In both cases, the top and middle panels compare the $z$-evolution of the STS intensity at its center (blue curve), with the analytically predicted intensity value from Section~\ref{sec:4} (dashed gray line).

For $E=6$, the evolution of the STS intensity is not constant, but it fluctuates around a value that is slightly larger than what is predicted by the analytical theory. We may understand this, if we remember that in the description of the effective dynamics the STS is a center equilibrium, and therefore, neutrally stable, which means that even numerical noise may perturb such equilibrium. In any case, simulations reveal the presence of fast, small amplitude intensity fluctuations as shown in Fig.~\ref{3D_simu_E6}(a),(b). This behavior agrees with that obtained when studying the effective reduced system (\ref{DS}) (see Sec.~\ref{sec:6.3}). These intensity fluctuations are depicted in more detail in Fig.~\ref{3D_simu_E6}(b) for the reduced interval $z\in (950,1000)$. 
It is worth noting that the weak fast fluctuations mainly result from the self-imaging effect due to the beating of transverse symmetric 
Laguerre-Gaussian modes \cite{Hansson2020}. They have a fixed period  
around $\pi/2$ which is slightly modified due to the phase accumulation from the second dispersion and the Kerr term.
The shape evolution of the STS in such an interval is illustrated in Fig.~\ref{3D_simu_E6}(c) by considering two isosurfaces at intensities $I_1=0.5$ and $I_2=0.1$, respectively. Figures~\ref{3D_simu_E6}(d)-(f) show that a similar evolution occurs for $E=8$. However, in this case, the amplitude of the STS intensity oscillations is larger than for the case with $E=6$.

The discrepancy between the analytically-obtained stable STS, and numerical results become larger when we increase STS energy. The center intensity of STS obtained from either the analytical method or from numerical simulations in the interval $z\in (0,1000)$ is compared in Fig.~\ref{fig_peak_vs_e}, by using a red line and blue circles, respectively. For numerical solutions, the blue circles and the error bars represent the time-averaged intensity values and the corresponding standard deviation for stable states.
For low values of $E$, the agreement is quite good, but it worsens with increasing $E$.  Eventually, for energy values above $E\approx 8.5$ the system undergoes full wave collapse. This region is illustrated by using a blue shadowed area. This collapse occurs much earlier than the analytical existence limit predicted by the theory at $E=E_f$ (see red dot). 
\begin{figure}[tbp]
\centering	\includegraphics[width=\columnwidth]{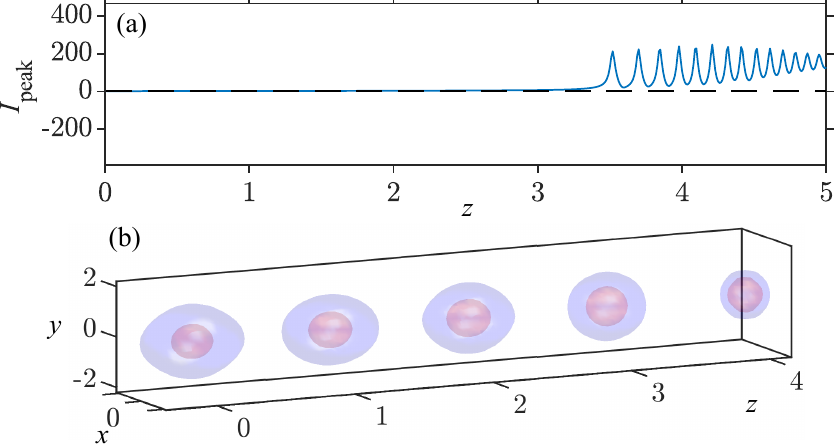}
\caption{Wave collapse started from an STS for $E=9$. Panel (a) shows the evolution of the $I_{peak}$, while panel (b) shows the modification of the STS with the propagation.  }
\label{collapsed}      
\end{figure}
An example of such destructive dynamics is illustrated in Fig.~\ref{collapsed}(a),(b) for $E=9$. The theoretically stable STS maintains its stability for a very short propagation length, but eventually, at $z\approx3.5$, it undergoes wave collapse. This is characterized by a very fast growth rate of the peak intensity [see Fig.~\ref{collapsed}(a)]. This concentration of the field intensity at the center of the state can also be observed in the STS shell evolution which is shown in Fig.~\ref{collapsed}(b).

So, although it is not predicted by the reduced effective dynamics description, the $z$-evolution of high energy STSs in the GPE suffers from wave collapse, owing to the self-focusing Kerr effect. 
The disagreement between theory and numerical results, which to our knowledge was not disclosed in earlier works, might be corrected by considering higher-order self-defocusing nonlinearities (e.g., quintic-order nonlinear terms), or high-order dispersive effects. However, these investigations will be presented elsewhere.

\begin{figure}[tbp]
\centering	\includegraphics[width=\columnwidth]
{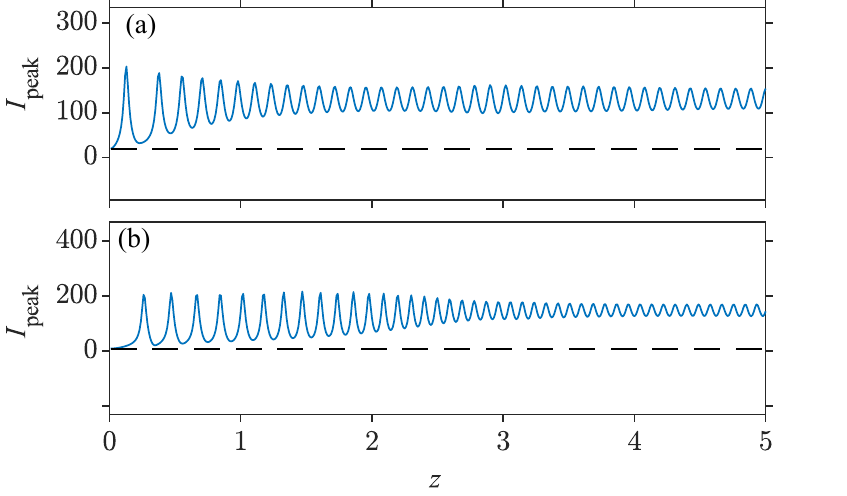}
\caption{(a) $z$-evolution of unstable STS solutions for $E=6$ showing wave collapse. (b) Shows similar dynamics for $E=8$. }
\label{Collapse_unstable}       
\end{figure}

We have performed similar simulations involving the analytically predicted unstable STSs for different values of $E$ [see Fig.~\ref{Collapse_unstable}(a) for $E=6$ and Fig.~\ref{Collapse_unstable}(b) for $E=8$]. According to the numerical simulation carried out on the reduced system (\ref{DS}), the unstable fixed point corresponding to this STS evolves into an oscillating continuous-wave state [see the evolution of $a$ and $\alpha$ in the right column of Fig.~\ref{fig5}], that we have identified with self-imaging \cite{karlsson_dynamics_1992}. However, here, 3D simulations initialized from the unstable STS variational solution do not converge to the self-imaging state, but rather undergo wave collapse almost immediately. This can be easily appreciated from the initial fast growth of $I_{peak}$, which is depicted in Fig.~\ref{Collapse_unstable}. 

\subsection{Normal/self-defocusing scenario} 
Similarly to the simulations performed in Sec.~\ref{sec:7.2}, here we test the deviation of the analytically computed variational STS solutions in the normal GVD/self-defocusing scenario, with respect to the exact numerical solutions which are obtained by directly solving Eq.~(\ref{GPE}). The comparison between the numerical and the approximate solutions is reported in Fig.~\ref{fig_peak_vs_e_normal}.  As it was described in Sec.~\ref{sec:7.2}, the stability of STSs is only confirmed for low values of $E$: the agreement is excellent for low values of $E$, but it becomes progressively worse when increasing the STS energy. However, despite these differences, in this regime, STS does not suffer wave collapse, but they undergo complex oscillatory dynamics, which may arise from secondary instabilities suffered by the STS. However, a more in-depth study of this scenario will require further investigations, which are beyond the scope of this work.

\section{Discussions and conclusions}\label{sec:8}
In this work, we have presented a complete and systematic analysis of 3D soliton solutions of the 3D+1 Gross-Pitaevskii equation (GPE) with a 2D parabolic potential. This equation can be used to describe light propagation in graded-index nonlinear media \cite{yu_spatio-temporal_1995,raghavan_spatiotemporal_2000,shtyrina_coexistence_2018} and for understanding the dynamics of nearly 1D condensates with a cigar-shape potential \cite{strecker_formation_2002}. In the nonlinear optics framework, our solutions are known as spatiotemporal solitons (STSs). The GPE with the 2D potential has a Lagrangian structure, which we have introduced in Section~\ref{sec:1} for the $z$-dependent and independent dynamics. Analytical approximations for soliton solutions can be computed through the Ritz optimization approach, by using an adequate parameter-dependent solution ansatz (see Section~\ref{sec:2}). This approach, based on the variational method, allows for reducing the GPE to a finite-dimensional dynamical system, associated with the evolution of the soliton parameters. In Section~\ref{sec:3}, by using the $z$-independent Lagrangian we computed the evolution of the parameters of a shape-preserving (i.e., steady state) STS, including its propagation constant $\mu$. This allowed us to understand the dependence of the STS solution upon its energy $E$, which we use as the main control parameter. In the anomalous GVD (self-focusing) regime, the STS existence region has an upper limit, corresponding to a fold point at $E_f$ in [see Fig.~\ref{fig1}(a)-(c)]. Below this point, two families of STSs $\mathcal{B}_a$ and $\mathcal{B}_b$ coexist. In the normal GVD (self-defocusing) regime, only one family of STS solutions exists, which persists for any value of $E$.

The previously described calculations can be generalized for $z$-dependent states. This can be done by considering the Ritz optimization method in either a Lagrangian or a Hamiltonian framework, as shown in Section~\ref{sec:4} and Section~\ref{sec:5}. In this case, the essential information on the effective dynamics of the STS solution is provided by the reduced 4D dynamical system (\ref{DS}). In Section~\ref{sec:6} we have used       
all of this characterization to determine the STSs stability by using three different approaches: the VKS criterion, which uses the $\mu$ vs. $E$ dependence (see Section~\ref{sec:6.1}); the Lyapunov stability criterion, which considers the $H$ vs. $E$ dependence (Section~\ref{sec:6.2}), and the spectral stability criterion (see Section~\ref{sec:6.3}), where stability is studied in terms of the linearization of Eq.~(\ref{DS}) around its equilibria (i.e., steady-state STS solutions). We find a perfect agreement between these three different criteria. Furthermore, by numerically solving Eq.~(\ref{DS}), we have studied the STSs permanent dynamics (Section~\ref{sec:6.3}).

\begin{figure}[tbp]
\centering	\includegraphics[width=\columnwidth]{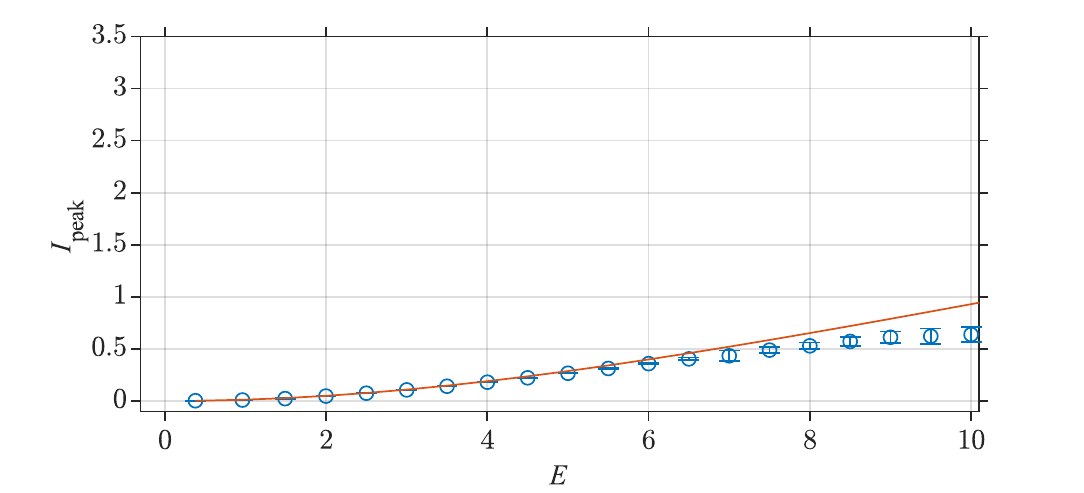}
\caption{Evolution of peak intensity of stable STS with energy $E$ in the normal GVD/self-defocusing regime (i.e., $\delta=\nu=-1$). The red line shows the analytical value which is obtained from the variational approach, while the blue dots represent the average intensity value obtained from full 3D numerical simulations.}
\label{fig_peak_vs_e_normal}      
\end{figure}

Finally, in Section~\ref{sec:7} we have tested our analytical predictions by performing extensive numerical solutions of the initial value problem associated with the full 3D+1 GPE (\ref{GPE}). By doing so, we demonstrated that, for low $E$, the agreement between variational approaches and numerical simulations is excellent, as depicted in Figs.~\ref{fig_peak_vs_e} and \ref{fig_peak_vs_e_normal}. Increasing $E$, however, a disagreement appears in both the anomalous and the normal GVD regime. In the first case, the STSs suffer wave collapse much below the theoretically predicted STS upper limit. Whereas in the second case, STSs undergo complex spatiotemporal dynamics, that will be analyzed in detail elsewhere. Such limitations of the variational approach were not investigated in previous works \cite{yu_spatio-temporal_1995,raghavan_spatiotemporal_2000,shtyrina_coexistence_2018}. 

In future works, we will explore different mechanisms which could be capable of stabilizing the observed STS instabilities. One of the possible paths to follow is to consider higher-order nonlinearities, which may come into play for very high values of $E$ \cite{karlsson_optical_1992,berezhiani_large_1995}. In particular, we will consider self-defocusing (self-focusing) quintic nonlinearities for the self-focusing (self-defocusing) Kerr nonlinear regimes that we have studied here. Note that quintic nonlinear effects have been considered in Ref.~\cite{skarka_spatiotemporal_1997}, but in the absence of the parabolic potential.
It would also be interesting to analyze how higher-order dispersion effects, such as fourth-order dispersion, may affect STS instabilities.   

Additional research could focus on other types of spatiotemporal potentials, such as Gaussian, tapered, or helicoidal \cite{strinic_light_2020}, and on studying the emergence of 3D spatiotemporal vortex solitons (which were extensively studied in BECs and other optical devices \cite{malomed_multidimensional_2016,veretenov_topological_2017,mayteevarunyoo_spatiotemporal_2019,kartashov_frontiers_2019,paredes_vortex_2022}). It would also be interesting to explore the connection between conservative STSs and the recently discovered dissipative STS in externally-driven multimode cavities \cite{sun_dissipative_2022}. 

\section*{Acknowledgements}
We are grateful to Tobias Hansson for helpful discussions and suggestions regarding the numerical simulations presented in Sec.~\ref{sec:7} . This work was supported by European Research Council (740355), Marie Sklodowska-Curie Actions (101064614,101023717), Ministero dell’Istruzione, dell’Università e della Ricerca (R18SPB8227).









\bibliographystyle{ieeetr}
\bibliography{References}

\begin{thebibliography}{10}

\bibitem{dauxois_physics_2006}
T.~Dauxois and M.~Peyrard, {\em Physics of {Solitons}}.
\newblock Cambridge University Press, Mar. 2006.
\newblock Google-Books-ID: YKe1UZc\_Qo8C.

\bibitem{PhysRevLett.15.240}
N.~J. Zabusky and M.~D. Kruskal, ``Interaction of "solitons" in a collisionless
  plasma and the recurrence of initial states,'' {\em Phys. Rev. Lett.},
  vol.~15, pp.~240--243, Aug 1965.

\bibitem{kartashov_frontiers_2019}
Y.~V. Kartashov, G.~E. Astrakharchik, B.~A. Malomed, and L.~Torner, ``Frontiers
  in multidimensional self-trapping of nonlinear fields and matter,'' {\em
  Nature Reviews Physics}, vol.~1, pp.~185--197, Mar. 2019.
\newblock Number: 3 Publisher: Nature Publishing Group.

\bibitem{malomed_multidimensional_nodate}
B.~A. Malomed, ``Multidimensional {Solitons},''
\newblock Publisher: AIP Publishing LLCAIP PublishingMelville, New York.

\bibitem{kivshar_optical_2003}
Y.~S. Kivshar, G.~P. Agrawal, and Y.~S. Kivshar, {\em Optical {Solitons}:
  {From} {Fibers} to {Photonic} {Crystals}}.
\newblock Mar. 2003.

\bibitem{silberberg_collapse_1990}
Y.~Silberberg, ``Collapse of optical pulses,'' {\em Optics Letters}, vol.~15,
  pp.~1282--1284, Nov. 1990.
\newblock Publisher: Optica Publishing Group.

\bibitem{berge_wave_1998}
L.~Bergé, ``Wave collapse in physics: principles and applications to light and
  plasma waves,'' {\em Physics Reports}, vol.~303, pp.~259--370, Sept. 1998.

\bibitem{bang_collapse_2002}
O.~Bang, W.~Krolikowski, J.~Wyller, and J.~J. Rasmussen, ``Collapse arrest and
  soliton stabilization in nonlocal nonlinear media,'' {\em Physical Review E},
  vol.~66, p.~046619, Oct. 2002.
\newblock Publisher: American Physical Society.

\bibitem{garmire_dynamics_1966}
E.~Garmire, R.~Y. Chiao, and C.~H. Townes, ``Dynamics and {Characteristics} of
  the {Self}-{Trapping} of {Intense} {Light} {Beams},'' {\em Physical Review
  Letters}, vol.~16, pp.~347--349, Feb. 1966.
\newblock Publisher: American Physical Society.

\bibitem{sackett_measurements_1999}
C.~A. Sackett, J.~M. Gerton, M.~Welling, and R.~G. Hulet, ``Measurements of
  {Collective} {Collapse} in a {Bose}-{Einstein} {Condensate} with {Attractive}
  {Interactions},'' {\em Physical Review Letters}, vol.~82, pp.~876--879, Feb.
  1999.
\newblock Publisher: American Physical Society.

\bibitem{wong_three-dimensional_1984}
A.~Y. Wong and P.~Y. Cheung, ``Three-{Dimensional} {Self}-{Collapse} of
  {Langmuir} {Waves},'' {\em Physical Review Letters}, vol.~52, pp.~1222--1225,
  Apr. 1984.
\newblock Publisher: American Physical Society.

\bibitem{noauthor_black_1983}
``Black {Holes},'' in {\em Black {Holes}, {White} {Dwarfs}, and {Neutron}
  {Stars}}, pp.~335--369, John Wiley \& Sons, Ltd, 1983.

\bibitem{malomed_multidimensional_2016}
B.~A. Malomed, ``Multidimensional solitons: {Well}-established results and
  novel findings,'' {\em The European Physical Journal Special Topics},
  vol.~225, pp.~2507--2532, Nov. 2016.

\bibitem{yu_spatio-temporal_1995}
S.-S. Yu, C.-H. Chien, Y.~Lai, and J.~Wang, ``Spatio-temporal solitary pulses
  in graded-index materials with {Kerr} nonlinearity,'' {\em Optics
  Communications}, vol.~119, pp.~167--170, Aug. 1995.

\bibitem{raghavan_spatiotemporal_2000}
S.~Raghavan and G.~P. Agrawal, ``Spatiotemporal solitons in inhomogeneous
  nonlinear media,'' {\em Optics Communications}, vol.~180, pp.~377--382, June
  2000.

\bibitem{horak_multimode_2012}
P.~Horak and F.~Poletti, ``Multimode {Nonlinear} {Fibre} {Optics}: {Theory} and
  {Applications},'' {\em Recent Progress in Optical Fiber Research}, Jan. 2012.

\bibitem{renninger_optical_2013}
W.~H. Renninger and F.~W. Wise, ``Optical solitons in graded-index multimode
  fibres,'' {\em Nature Communications}, vol.~4, p.~1719, Apr. 2013.

\bibitem{panagiotopoulos_super_2015}
P.~Panagiotopoulos, P.~Whalen, M.~Kolesik, and J.~V. Moloney, ``Super high
  power mid-infrared femtosecond light bullet,'' {\em Nature Photonics},
  vol.~9, pp.~543--548, Aug. 2015.
\newblock Number: 8 Publisher: Nature Publishing Group.

\bibitem{minardi_three-dimensional_2010}
S.~Minardi, F.~Eilenberger, Y.~V. Kartashov, A.~Szameit, U.~Röpke, J.~Kobelke,
  K.~Schuster, H.~Bartelt, S.~Nolte, L.~Torner, F.~Lederer, A.~Tünnermann, and
  T.~Pertsch, ``Three-{Dimensional} {Light} {Bullets} in {Arrays} of
  {Waveguides},'' {\em Physical Review Letters}, vol.~105, p.~263901, Dec.
  2010.
\newblock Publisher: American Physical Society.

\bibitem{shtyrina_coexistence_2018}
O.~V. Shtyrina, M.~P. Fedoruk, Y.~S. Kivshar, and S.~K. Turitsyn, ``Coexistence
  of collapse and stable spatiotemporal solitons in multimode fibers,'' {\em
  Physical Review A}, vol.~97, p.~013841, Jan. 2018.
\newblock Publisher: American Physical Society.

\bibitem{strecker_formation_2002}
K.~E. Strecker, G.~B. Partridge, A.~G. Truscott, and R.~G. Hulet, ``Formation
  and propagation of matter-wave soliton trains,'' {\em Nature}, vol.~417,
  pp.~150--153, May 2002.
\newblock Number: 6885 Publisher: Nature Publishing Group.

\bibitem{abraham_foundations_2008}
R.~Abraham and J.~E. Marsden, {\em Foundations of {Mechanics}}.
\newblock American Mathematical Soc., 2008.

\bibitem{wiggins_introduction_2003}
S.~Wiggins, {\em Introduction to {Applied} {Nonlinear} {Dynamical} {Systems}
  and {Chaos}}.
\newblock Texts in {Applied} {Mathematics}, New York: Springer-Verlag, 2~ed.,
  2003.

\bibitem{anderson_selftrapped_1979}
D.~Anderson, M.~Bonnedal, and M.~Lisak, ``Self‐trapped cylindrical laser
  beams,'' {\em The Physics of Fluids}, vol.~22, pp.~1838--1840, Sept. 1979.
\newblock Publisher: American Institute of Physics.

\bibitem{bondeson_soliton_1979}
A.~Bondeson, M.~Lisak, and D.~Anderson, ``Soliton {Perturbations}: {A}
  {Variational} {Principle} for the {Soliton} {Parameters},'' {\em Physica
  Scripta}, vol.~20, p.~479, Sept. 1979.

\bibitem{perez-garcia_dynamics_1997}
V.~M. Pérez-García, H.~Michinel, J.~I. Cirac, M.~Lewenstein, and P.~Zoller,
  ``Dynamics of {Bose}-{Einstein} condensates: {Variational} solutions of the
  {Gross}-{Pitaevskii} equations,'' {\em Physical Review A}, vol.~56,
  pp.~1424--1432, Aug. 1997.

\bibitem{malomed_variational_2002}
B.~A. Malomed, ``Variational methods in nonlinear fiber optics and related
  fields,'' in {\em Progress in {Optics}}, vol.~43, pp.~71--193, Elsevier,
  2002.

\bibitem{rasmussen_blow-up_1986}
J.~J. Rasmussen and K.~Rypdal, ``Blow-up in {Nonlinear} {Schroedinger}
  {Equations}-{I} {A} {General} {Review},'' {\em Physica Scripta}, vol.~33,
  p.~481, June 1986.

\bibitem{montesinos_stabilization_2004}
G.~Montesinos, ``Stabilization of solitons of the multidimensional nonlinear
  {Schrödinger} equation: matter-wave breathers,'' {\em Physica D: Nonlinear
  Phenomena}, vol.~191, pp.~193--210, May 2004.

\bibitem{hansson_nonlinear_2020}
T.~Hansson, A.~Tonello, T.~Mansuryan, F.~Mangini, M.~Zitelli, M.~Ferraro,
  A.~Niang, R.~Crescenzi, S.~Wabnitz, and V.~Couderc, ``Nonlinear beam
  self-imaging and self-focusing dynamics in a {GRIN} multimode optical fiber:
  theory and experiments,'' {\em Optics Express}, vol.~28, pp.~24005--24021,
  Aug. 2020.
\newblock Publisher: Optica Publishing Group.

\bibitem{vakhitov_stationary_1973}
N.~G. Vakhitov and A.~A. Kolokolov, ``Stationary solutions of the wave equation
  in a medium with nonlinearity saturation,'' {\em Radiophysics and Quantum
  Electronics}, vol.~16, pp.~783--789, July 1973.

\bibitem{sakaguchi_two-dimensional_2006}
H.~Sakaguchi and B.~A. Malomed, ``Two-dimensional solitons in the
  {Gross}-{Pitaevskii} equation with spatially modulated nonlinearity,'' {\em
  Physical Review E}, vol.~73, p.~026601, Feb. 2006.
\newblock Publisher: American Physical Society.

\bibitem{malomed_stability_2007}
B.~A. Malomed, F.~Lederer, D.~Mazilu, and D.~Mihalache, ``On stability of
  vortices in three-dimensional self-attractive {Bose}–{Einstein}
  condensates,'' {\em Physics Letters A}, vol.~361, pp.~336--340, Feb. 2007.

\bibitem{desyatnikov_three-dimensional_2000}
A.~Desyatnikov, A.~Maimistov, and B.~Malomed, ``Three-dimensional spinning
  solitons in dispersive media with the cubic-quintic nonlinearity,'' {\em
  Physical Review E}, vol.~61, pp.~3107--3113, Mar. 2000.
\newblock Publisher: American Physical Society.

\bibitem{baizakov_multidimensional_2004}
B.~B. Baizakov, B.~A. Malomed, and M.~Salerno, ``Multidimensional solitons in a
  low-dimensional periodic potential,'' {\em Physical Review A}, vol.~70,
  p.~053613, Nov. 2004.
\newblock Publisher: American Physical Society.

\bibitem{desaix_variational_1991}
M.~Desaix, D.~Anderson, and M.~Lisak, ``Variational approach to collapse of
  optical pulses,'' {\em Journal of the Optical Society of America B}, vol.~8,
  p.~2082, Oct. 1991.

\bibitem{skarka_spatiotemporal_1997}
V.~Skarka, V.~I. Berezhiani, and R.~Miklaszewski, ``Spatiotemporal soliton
  propagation in saturating nonlinear optical media,'' {\em Physical Review E},
  vol.~56, pp.~1080--1087, July 1997.
\newblock Publisher: American Physical Society.

\bibitem{skarka_stability_2006}
V.~Skarka and N.~B. Aleksić, ``Stability {Criterion} for {Dissipative}
  {Soliton} {Solutions} of the {One}-, {Two}-, and {Three}-{Dimensional}
  {Complex} {Cubic}-{Quintic} {Ginzburg}-{Landau} {Equations},'' {\em Physical
  Review Letters}, vol.~96, p.~013903, Jan. 2006.

\bibitem{reiszig_j_1962}
R.~Reiszig, ``J. {LaSalle} and {S}. {Lefschetz}, {Stability} by {Liapunov}'s
  {Direct} {Method} with {Applications}. {VII} + 134 {S}. {New} {York}/{London}
  1961. {Academic} {Press}. {Preis} geb. \$ 5.50,'' {\em ZAMM - Journal of
  Applied Mathematics and Mechanics / Zeitschrift für Angewandte Mathematik
  und Mechanik}, vol.~42, no.~10-11, pp.~514--514, 1962.

\bibitem{rosen_particlelike_1965}
G.~Rosen, ``Particlelike {Solutions} to {Nonlinear} {Scalar} {Wave}
  {Theories},'' {\em Journal of Mathematical Physics}, vol.~6, pp.~1269--1272,
  Aug. 1965.
\newblock Publisher: American Institute of Physics.

\bibitem{akhmediev_hamiltonian-versus-energy_1999}
N.~Akhmediev, A.~Ankiewicz, and R.~Grimshaw, ``Hamiltonian-versus-energy
  diagrams in soliton theory,'' {\em Physical Review E}, vol.~59,
  pp.~6088--6096, May 1999.
\newblock Publisher: American Physical Society.

\bibitem{karlsson_dynamics_1992}
M.~Karlsson, D.~Anderson, and M.~Desaix, ``Dynamics of self-focusing and
  self-phase modulation in a parabolic index optical fiber,'' {\em Optics
  Letters}, vol.~17, p.~22, Jan. 1992.

\bibitem{agrawal_applications_2008}
G.~Agrawal, {\em Applications of {Nonlinear} {Fiber} {Optics}}.
\newblock Academic Press, Mar. 2008.
\newblock Google-Books-ID: HbkKQLPE8yEC.

\bibitem{frolkovic_numerical_1990}
P.~Frolkovič, ``Numerical recipes: {The} art of scientific computing,'' {\em
  Acta Applicandae Mathematica}, vol.~19, pp.~297--299, June 1990.

\bibitem{Hansson2020}
T.~Hansson, A.~Tonello, T.~Mansuryan, F.~Mangini, M.~Zitelli, M.~Ferraro,
  A.~Niang, R.~Crescenzi, S.~Wabnitz, and V.~Couderc, ``Nonlinear beam
  self-imaging and self-focusing dynamics in a grin multimode optical fiber:
  theory and experiments,'' {\em Opt. Express}, vol.~28, p.~24005, 2020.

\bibitem{karlsson_optical_1992}
M.~Karlsson, ``Optical beams in saturable self-focusing media,'' {\em Physical
  Review A}, vol.~46, pp.~2726--2734, Sept. 1992.
\newblock Publisher: American Physical Society.

\bibitem{berezhiani_large_1995}
V.~I. Berezhiani and S.~M. Mahajan, ``Large relativistic density pulses in
  electron-positron-ion plasmas,'' {\em Physical Review E}, vol.~52,
  pp.~1968--1979, Aug. 1995.
\newblock Publisher: American Physical Society.

\bibitem{strinic_light_2020}
A.~I. Strinić, N.~B. Aleksić, M.~R. Belić, and M.~S. Petrović, ``Light
  propagation along a helical waveguide: variational approach,'' {\em Optical
  and Quantum Electronics}, vol.~52, p.~310, June 2020.

\bibitem{veretenov_topological_2017}
N.~Veretenov, S.~Fedorov, and N.~Rosanov, ``Topological {Vortex} and {Knotted}
  {Dissipative} {Optical} {3D} {Solitons} {Generated} by {2D} {Vortex}
  {Solitons},'' {\em Physical Review Letters}, vol.~119, p.~263901, Dec. 2017.
\newblock Publisher: American Physical Society.

\bibitem{mayteevarunyoo_spatiotemporal_2019}
T.~Mayteevarunyoo, B.~A. Malomed, and D.~V. Skryabin, ``Spatiotemporal
  dissipative solitons and vortices in a multi-transverse-mode fiber laser,''
  {\em Optics Express}, vol.~27, pp.~37364--37373, Dec. 2019.

\bibitem{paredes_vortex_2022}
A.~Paredes, J.~R. Salgueiro, and H.~Michinel, ``On vortex and dark solitons in
  the cubic–quintic nonlinear {Schrödinger} equation,'' {\em Physica D:
  Nonlinear Phenomena}, vol.~437, p.~133340, Sept. 2022.

\bibitem{sun_dissipative_2022}
Y.~Sun, P.~Parra-Rivas, M.~Ferraro, F.~Mangini, M.~Zitelli, R.~Jauberteau,
  F.~R. Talenti, and S.~Wabnitz, ``Dissipative {Kerr} solitons, breathers and
  chimera states in coherently driven passive cavities with parabolic
  potential,'' Aug. 2022.
\newblock arXiv:2208.12669 [nlin, physics:physics].

\end{thebibliography}

\end{document}